\documentclass[letterpaper]{JHEP3}
\usepackage{amsfonts}

\newcommand{\vphi}{\varphi}
\usepackage{epsfig}

\abstract{
We present nonuniform vacuum black strings in five and six spacetime dimensions.
The conserved charges and the action of these solutions are computed by employing
a quasilocal formalism.
We find qualitative agreement of the physical properties
of nonuniform black strings in five and six dimensions.
Our results offer further evidence that the black hole and the black string
branches merge at a topology changing transition.
We generate black string solutions
of the Einstein-Maxwell-dilaton theory by using a Harrison transformation. 
We argue that the basic features of these solutions can be derived from 
those of the vacuum black string configurations.
}

\title{New nonuniform black string solutions}
\author{Burkhard Kleihaus$^{1}$\thanks{%
E-mail: \texttt{kleihaus@theorie.physik.uni-oldenburg.de}}, \ Jutta Kunz$^{1}$%
\thanks{%
E-mail: \texttt{kunz@theorie.physik.uni-oldenburg.de}} \ and Eugen Radu$^2$\thanks{%
E-mail: \texttt{radu@thphys.nuim.ie}} \\
$^{1}$Institut f\"ur Physik, Universit\"at Oldenburg, Postfach 2503
D-26111 Oldenburg, Germany\\ 
$^{2} $Department of Mathematical Physics, National University of Ireland,
Maynooth, Ireland}

\begin{document}

%%%%%%%%%%%%%%%%%%%%%%%%%%%%%%%%%%%%%%%%%%%%%%%%%%%%%%
\section{Introduction}  
%%%%%%%%%%%%%%%%%%%%%%%%%%%%%%%%%%%%%%%%%%%%%%%%%%%%%%
In recent years it has been realized that, 
even at the classical level, 
gravity exhibits a much richer structure in higher dimensions
than in four dimensions.

Black string solutions, present for
$D\geq 5$ spacetime dimensions, are of particular interest, since they 
exhibit new features that have no analogue in the black hole case.
Such configurations are important if one supposes 
the existence of extradimensions in the universe, 
which are likely to be compact and described by a Kaluza-Klein (KK) theory.

The simplest vacuum static solution of this 
type 
is found by assuming translational symmetry along the 
extracoordinate direction, and corresponds to a uniform black string
with horizon topology $S^{D-3}\times S^1$. 
Although this solution exists for all values of the mass, 
it is unstable below a critical value
as shown by Gregory and Laflamme \cite{Gregory:1993vy}.
This instability was interpreted to mean that a light uniform string decays
to a  black hole since that has higher entropy.
However, Horowitz and
Maeda~\cite{Horowitz:2001cz} argued that the horizon could not pinch off, so the
end state of the instability could not be a collection of separate
black holes. Instead, they conjectured that the solution would settle
down to a non-translationally invariant solution with the same horizon
topology as the original configuration. 

This prompted a search for this missing link, and a branch of
static nonuniform black string solutions was subsequently found by
perturbing the uniform black string by the threshold unstable
mode. The $D=5$ approximate   solutions were obtained in 
\cite{Gubser:2001ac}, perturbed  $D=6$ were presented in  \cite{Wiseman:2002zc},
and higher
dimensional generalizations were discussed in \cite{Sorkin:2004qq}.

Despite some attempts,
no analytic solutions are available for nonuniform black string solutions
and one has to employ  numerical techniques.
At present, the nonuniform branch is numerically known 
only for $D=6$ \cite{Wiseman:2002zc} (see also the post-analysis in 
\cite{Wiseman:2002ti}, \cite{Kol:2003ja}).

Apart from the black string solutions, KK theory possesses also a 
branch of black hole solutions with an event horizon of topology $S^{D-2}$.
The numerical results presented in \cite{Kudoh:2004hs} 
(following a conjecture put forward in \cite{Kol:2002xz})
suggest that, at least for $D=6$, the black hole and the nonuniform
string branches merge at a topology changing transition.
Still, a number of aspects remain to be clarified,
and 
the literature on nonuniform black string and black hole solutions 
in KK theory is continuously growing 
(see \cite{Kol:2004ww,Harmark:2005pp} for recent reviews).
 
The main purpose of this paper is to numerically construct and study the nonuniform 
black string branch in $D=5$.
This dimension is of particular interest since 
one may join the black string results with those of the $D=5$ black hole 
branch discussed in  \cite{Kudoh:2004hs} (see also \cite{Sorkin:2003ka}).

We begin with a presentation of the general ansatz and the relevant quantities for an arbitrary spacetime
dimension $D$.
In this context, we propose to compute
the mass, tension and action of the nonuniform black string solutions  
by using a quasilocal formalism.
In Section 3 we present our numerical results. 
We demonstrate that for $D=5$
a branch of  nonuniform black string solutions exists,
at least within the scope of our numerical approximation.
The numerical methods used here are rather different from the methods 
employed to obtain the $D=6$ solutions \cite{Wiseman:2002zc}.
We construct nonuniform black string solutions
also in $D=6$ dimensions, %reproducing and 
extending the known set of solutions \cite{Wiseman:2002zc} to 
larger deformation of the event horizon.
In Section 4 solutions of the Einstein-Maxwell-dilaton (EMD) equations are generated
from the vacuum configurations, by using a Harrison transformation
originally derived in \cite{Gal'tsov:1998yu}.
The basic properties of two different types of solutions, corresponding to charged 
black strings which asymptote
to ${\cal M}^{D-1}\times S^1$ and black strings in a background magnetic field are discussed there.
We give our conclusions and remarks in the final section.
The Appendix contains a brief discussion of some numerical aspects of our solutions.

%%%%%%%%%%%%%%%%%%%%%%%%%%%%%%%%%%%%%%%%%%%%%%%%%%%%%%
\section{General ansatz and properties of the solutions}  
%%%%%%%%%%%%%%%%%%%%%%%%%%%%%%%%%%%%%%%%%%%%%%%%%%%%%%

\subsection{The equations and boundary conditions}  
We consider the Einstein action
\begin{eqnarray} 
\label{action-grav} 
I=\frac{1}{16 \pi G}\int_M~d^Dx \sqrt{-g} R
-\frac{1}{8\pi G}\int_{\partial M} d^{D-1}x\sqrt{-h}K,
\end{eqnarray}
in a $D-$dimensional spacetime.
The last term in  (\ref{action-grav}) is the Hawking-Gibbons surface term \cite{Gibbons:1976ue},
which
is required in order to have a well-defined variational principle. 
$K$ is the trace 
of the extrinsic curvature for the boundary $\partial\mathcal{M}$ and $h$ is the induced 
metric of the boundary.  
 
We consider black string solutions approaching asymptotically the 
$D-1$ dimensional Minkowski-space times a circle ${\cal M}^{D-1}\times S^1$.
We denote the compact direction as $z = x^{D-1}$ 
and the directions of $R^{D-2}$ as $x^1,...,x^{D-2}$, while $x^D=t$.
The direction $z$ is periodic with period $L$.
We also define the radial coordinate $r$ by
$r^2 = (x^1)^2 + \cdots + (x^{D-2})^2$.

The nonuniform black string solutions presented in this paper 
are found within the metric ansatz  
\footnote{
An ansatz involving only two undetermined functions
was suggested in \cite{Harmark:2002tr} 
for a special coordinate choice.
However, we could not obtain numerical solutions of
the Einstein equations within that reduced ansatz.}
\begin{eqnarray} 
\label{metric}
ds^2=-e^{2A(r,z)}f(r)dt^2+e^{2B(r,z)}\left(\frac{dr^2}{f(r)}+dz^2\right)+e^{2C(r,z)}r^2d\Omega_{D-3}^2
, \end{eqnarray}
where
\begin{eqnarray}
\nonumber 
f=1-(\frac{r_0}{r})^{D-4}  .
\end{eqnarray}

The Einstein equations $G_t^t=0,~G_r^r+G_z^z=0$ and $G_\theta^\theta=0$ 
(where $\theta$ denotes an angle of the $D-3$ dimensional sphere) 
then yield for the functions $A,~B,~C$ the set of equations
\cite{Wiseman:2002zc}
\begin{eqnarray} 
\nonumber
&&A''+\frac{{\ddot A}}{f}+A'^2+\frac{{\dot A}^2}{f}
+(D-3)\left(A'C'+\frac{{\dot A}{\dot C}}{f}+\frac{A'}{r}+\frac{f'C'}{2f}+\frac{f'}{2rf} \right)
+\frac{f''}{2f}+\frac{3f'A'}{2f}=0,
\\
\label{Eeq}
&&B''+\frac{{\ddot B}}{f}+\frac{(D-3)(D-4)}{2r^2}\left(-1+\frac{e^{2B-2C}}{f}
-\frac{r^2{\dot C}^2}{f}-2rC'-r^2C'^2\right)
\\
\nonumber
&&{\rm ~~~~~~~~~~~~~~~~~~~~~~~~~~~~~~~~~~~~~~~~}
-(D-3)\left(\frac{f'}{2rf}+\frac{{\dot A}{\dot C}}{f}+\frac{A'}{r}
+\frac{f'C'}{2f}+A'C'\right)+\frac{f'B'}{2f}=0,
\\
\nonumber
&&C''+\frac{{\ddot C}}{f}+(D-3)
\left(
C'^2+\frac{{\dot C}^2}{f}+\frac{2C'}{r}
\right)
+\frac{(D-4)}{r^2}(1-\frac{e^{2B-2C}}{f})
\\
\nonumber
&&{\rm ~~~~~~~~~~~~~~~~~~~~~~~~~~~~~~~~~~~~~~~~~~~~~~~~~~~~~}
+\frac{f'}{rf}+\frac{{\dot A}{\dot C}+ f'C'}{f} +\frac{A'}{r}+A'C'=0,
\end{eqnarray}
where a prime denotes $\partial/\partial r$, and a 
dot $\partial/\partial z$.

The remaining Einstein equations $G_z^r=0,~G_r^r-G_z^z=0$
yield two constraints. Following \cite{Wiseman:2002zc}, we note that
setting $G^t_t=G^\theta_\theta=G^\vphi_\vphi=G^r_r+G^z_z=0$
in $\nabla_\mu G^{\mu r}=0$ and $\nabla_\mu G^{\mu z}=0$, we obtain
\begin{eqnarray}
\partial_z\left(\sqrt{-g} G^r_z \right) +
\sqrt{f} \partial_r\left( \sqrt{f}\sqrt{-g} \frac{1}{2}(G^r_r-G^z_z) \right)
& = & 0 ,
\\
\nonumber 
\sqrt{f}\partial_r\left(\sqrt{-g} G^r_z \right)
-\partial_z\left( \sqrt{f}\sqrt{-g} \frac{1}{2}(G^r_r-G^z_z) \right)
& = & 0 ,
\end{eqnarray}
and, defining $\hat{r}$ via
$\partial/\partial_{\hat{r}} = \sqrt{f}\partial/\partial_{r}$,
then yields the Cauchy-Riemann relations
\begin{eqnarray}
\partial_z\left(\sqrt{-g} G^r_z \right) +
\partial_{\hat{r}}\left( \sqrt{f}\sqrt{-g} \frac{1}{2}(G^r_r-G^z_z) \right)
& = & 0 ,\\
\nonumber 
\partial_{\hat{r}}\left(\sqrt{-g} G^r_z \right)
-\partial_z\left( \sqrt{f}\sqrt{-g} \frac{1}{2}(G^r_r-G^z_z) \right)
& = & 0  .
\end{eqnarray}
Thus the weighted constraints satisfy Laplace equations,
and the constraints are fulfilled,
when one of them is satisfied on the boundary 
and the other at a single point
\cite{Wiseman:2002zc}. 

The event horizon resides at a surface of constant radial coordinate
$r=r_0$ and is characterized by the condition $f(r_0)=0$.
Introducing the coordinate $\tilde r$, where $r=\sqrt{r_0^2 + \tilde r^2}$,
the horizon resides at $\tilde r=0$.

%%%%%%%%%%%%%%%%%%%%%%%%%%%%%%%%%%%%%%%%%%%%%%%%%%%%
%\subsection{Boundary conditions}
%%%%%%%%%%%%%%%%%%%%%%%%%%%%%%%%%%%%%%%%%%%%%%%%%%%%

Utilizing the reflection symmetry of the nonuniform black strings 
w.r.t.~$z=L/2$,
the solutions are constructed subject to the following set of 
boundary conditions
\begin{eqnarray}
\label{bc1} 
A\big|_{\tilde r=\infty}=B\big|_{\tilde r=\infty}=C\big|_{\tilde r=\infty}=0,
\end{eqnarray}
\begin{eqnarray}
\label{bc2} 
A\big|_{\tilde r=0}-B\big|_{\tilde r=0}=d_0,~\partial_{\tilde r} 
A\big|_{\tilde r=0}=\partial_{\tilde r} C\big|_{\tilde r=0}=0,
\end{eqnarray}
\begin{eqnarray}
\label{bc3} 
\partial_z A\big|_{z=0,L/2}=\partial_z B\big|_{z=0,L/2}
=\partial_z C\big|_{z=0,L/2}=0,
\end{eqnarray}
where the constant $d_0$ is related to the Hawking 
temperature of the solutions.
Regularity further requires that the condition 
$\partial_{\tilde r} B\big|_{\tilde r=0}=0$ holds for the solutions.
The boundary conditions guarantee, that the constraints are satisfied,
since $\sqrt{-g} G^r_z=0$ everywhere on the boundary,
and $\sqrt{f}\sqrt{-g}( G^r_r-G^z_z ) =0$ on the horizon.

%%%%%%%%%%%%%%%%%%%%%%%%%%%%%%%%%%%%%%%%%%%%%%%%%%%%
\subsection{Properties of the solutions}
%%%%%%%%%%%%%%%%%%%%%%%%%%%%%%%%%%%%%%%%%%%%%%%%%%%%

For any static spacetime which is asymptotically 
${\cal M}^{D-1}\times S^1$
one can define a mass  $M$ and a  tension ${\mathcal T}$ \cite{Traschen:2001pb},
these quantities being encoded in the asymptotics of the metric potentials.
As discussed in \cite{Harmark:2003dg}, \cite{Kol:2003if}, 
the asymptotic form of the  relevant metric components 
of any static solution is
\begin{eqnarray}
\label{1} 
g_{tt}\simeq -1+\frac{c_t}{r^{D-4}},~~~g_{zz}\simeq 1+\frac{c_z}{r^{D-4}}.
\end{eqnarray}
When computing $M$, ${\mathcal T}$ or the gravitational action, 
the essential idea is to consider the asymptotic
values of the gravitational field far away from the black string
and to compare them with those corresponding to a gravitational field
in the absence of the black string.
Therefore, this prescription provides results that are relative to the choice 
of a reference background, whose obvious choice in our case is ${\cal M}^{D-1}\times S^1$.

The mass and tension of black string solutions as computed in
\cite{Harmark:2003dg,Harmark:2004ch} are given by
\begin{eqnarray}
\label{2} 
M=\frac{\Omega_{D-3}L}{16 \pi G}((D-3)c_t-c_z),
~~{\mathcal T}=\frac{\Omega_{D-3}}{16 \pi G}(c_t-(D-3)c_z),
\end{eqnarray}
where $\Omega_{D-3}$ is the area of the unit $S^{D-3}$ sphere.
The corresponding quantities of the uniform string solution
$M_0$ and ${\mathcal T}_0$ are obtained from (\ref{2}) for $c_z=0$,
$c_t=r_0^{D-4}$.
One can also define a relative tension $n$ 
(also called the relative binding energy or scalar charge) 
\begin{eqnarray}
\label{3}
n=\frac{{\mathcal T} L}{M}=\frac{c_t-(D-3)c_z}{(D-3)c_t-c_z}.
\end{eqnarray}
which measures how large the tension is relative to the mass.
This dimensionless quantity is bounded, $0\leq n\leq D-3$.
Uniform string solutions have relative tension $n_0=1/(D-3)$.
Another useful quantity is the rescaled dimensionless mass
\begin{eqnarray}
\label{4}
\mu = \frac{16 \pi G M}{L^{D-3}}.
\end{eqnarray}
The Hawking temperature and entropy of the black string solutions are given by
\begin{eqnarray}
\label{temp} 
T=e^{A_0-B_0}T_0,~~~S=S_0\frac{1}{L}\int_0^L e^{B_0+(D-3)C_0}dz,
\end{eqnarray}
where $T_0,~S_0$ are the corresponding quantities of the uniform solution
\begin{eqnarray}
\label{temp0} 
T_0=\frac{D-4}{4 \pi r_0},~~~S_0= \frac{1}{4}L\Omega_{D-3}r_0^{D-3},
\end{eqnarray}
and $A_0(z),B_0(z),C_0(z)$ are the values of the metric functions on the event horizon $r=r_0$.

Together with the mass $M$ and relative tension $n$,
these quantities obey the Smarr formula \cite{Harmark:2003dg}
\begin{eqnarray}
\label{smarrform} 
TS = \frac{D-3-n}{D-2} M.
\end{eqnarray}
Note that the relations (\ref{1})-(\ref{4}) and (\ref{smarrform}) 
are also valid for black hole solutions.

Black string thermodynamics can be  discussed
by employing the very general 
connection between entropy and geometry 
established in the Euclidean path integral 
approach to quantum gravity \cite{Gibbons:1976ue}. 
In this approach, the  
partition function for the gravitational field is defined by a sum
over all smooth Euclidean geometries which are periodic with a period
$\beta$ in imaginary time.
This integral is computed by using the saddle point approaximation,
and the energy and entropy of the solutions are evaluated 
by standard thermodynamic formulae.

We consider a canonical ensemble with Helmholz free energy
(thus at fixed temperature and extradimension length)
\begin{eqnarray} 
F[T,L]=\frac{I}{\beta}=M-TS
\end{eqnarray}
The Euclidean action of the vacuum solutions  computed by subtracting the 
 background contribution is
\begin{eqnarray} 
I=\frac{1}{16 \pi G} \beta \Omega_{D-3} (c_t-c_z)=\frac{\beta}{D-2}(M+{\mathcal T}L),
\end{eqnarray}
with $\beta=1/T$.

The first law of thermodynamics reads
\begin{eqnarray}
\label{firstlaw}
dM=TdS+{\mathcal T}dL. 
\end{eqnarray}
It follows that
\begin{eqnarray} 
S=-\left(\frac{\partial F}{\partial T}\right)_{L},~~
{\mathcal T}=-\left(\frac{\partial F}{\partial L}\right)_{T}.
\end{eqnarray}
Combining the Smarr formula (\ref{smarrform}) and the first law, it follows that, given a curve 
$n(\mu)$ in the $(n,\mu)$-plane, 
the entire thermodynamics can be obtained \cite{Harmark:2003dg}.

We remark also that the Einstein equations (\ref{Eeq}) are left invariant
by the transformation $r \to k r$, $ z \to k z$, $ r_0 \to
r_0/k$, with $k$ an arbitrary positive integer.
Therefore, one may generate a family of vacuum solutions
in this way, termed copies of solutions.
The new solutions have the same length of the extradimension.
Their relevant properties, expressed in terms of the
corresponding properties of the initial solution, read
\begin{eqnarray}
M_k=\frac{M}{k^{D-4}}, ~~T_k=k T,~~S_k=\frac{S}{k^{D-3}},~~
n_k=n.
\end{eqnarray}
This transformation, suggested first by Horowitz \cite{Horowitz:2002dc},
has been discussed in \cite{Harmark:2003eg} for $D=6$.

%%%%%%%%%%%%%%%%%%%%%%%%%%%%%%%%%%%%%%%%%%%%%%%%%%%%
\subsection{A counterterm approach}
%%%%%%%%%%%%%%%%%%%%%%%%%%%%%%%%%%%%%%%%%%%%%%%%%%%%

Similar results for the black strings' mass, tension and action
are obtained by using the quasilocal
tensor of Brown and York \cite{Brown:1992br}, 
augmented by the counterterms formalism.
This technique consists 
in adding suitable counterterms $I_{ct}$
to the action of the theory.
These counterterms are built up with
curvature invariants of  the induced metric on the boundary $\partial \cal{M}$ 
(which is sent to infinity after the integration)
and thus obviously they do not alter the bulk equations of motion.
 By choosing 
appropriate counterterms which cancel the divergencies, one can then obtain 
well-defined expressions for the action and the energy momentum of the 
spacetime. Unlike the background substraction, this procedure is 
satisfying since it is intrinsic to the spacetime of interest  and it is 
unambiguous once the counterterm is specified. While there is a 
general algorithm to generate the counterterms for asymptotically (anti-)de Sitter 
spacetimes, the asymptotically flat case is 
less-explored (see however \cite{Kraus:1999di}
and the more general approach in recent work
\cite{Mann:2005yr}).

It is also important to note that the counterterm method  gives results
that are equivalent to what one obtains using the background subtraction method.
However, we employ it because it appears to be a more general technique than
background subtraction, and it is interesting to explore the range of problems
to which it applies. 

Therefore, 
we add  the following counterterm part to the action principle (\ref{action-grav})
\begin{eqnarray}
\label{ct}
I_{ct}=
-\frac{1}{8 \pi G}\int_{\partial\mathcal{M}} d^{D-1} x\sqrt{-h}
\sqrt{\frac{D-3}{D-4}\mathcal{R}},
\end{eqnarray} 
where  
$\mathcal{R}$ is the Ricci scalar of the boundary geometry.

Varying the total action with respect to the
boundary metric $h_{ij}$, we compute the  boundary stress-tensor
\begin{eqnarray}
\label{Tik}
T_{ij}=\frac{2}{\sqrt{-h}}\frac{\delta I}{\delta h^{ij}}=
\frac{1}{8\pi G}\Big( K_{ij}-h_{ij}K
-\Psi(\mathcal{R}_{ij}-\mathcal{R}h_{ij})-h_{ij}\Box \Psi+\Psi_{;ij}
\Big),
\end{eqnarray}
where $K_{ij}$ is the extrinsic curvature of the boundary and 
$\Psi=\sqrt{\frac{D-3}{(D-4)\mathcal{R}}}$.

Provided the boundary geometry has an isometry generated by a
Killing vector $\xi ^{i}$, a conserved charge
\begin{eqnarray}
{\cal Q}_{\xi }=\oint_{\Sigma }d^{D-2}S^{i}~\xi^{j}T_{ij}
\label{charge}
\end{eqnarray}
can be associated with a closed surface $\Sigma $.
Physically, 
this means that a collection of observers on the boundary with the 
induced metric $h_{ij}$ measure the same value of ${\cal Q}_{\xi }$.

The mass and tension are the charges associated to $\partial/\partial t$
and $\partial/\partial z$, respectively (note that $\partial/\partial z$
is a Killing symmetry of the boundary metric).
The relevant components of the boundary stress tensor are 
\begin{eqnarray} 
T_t^t=\frac{1}{16 \pi G}\frac{(D-3)c_t- c_z}{r^{D-3}}+O(1/r^{D-2}),~
T_z^z=\frac{1}{16 \pi G}\frac{c_t-(D-3)c_z}{r^{D-3}}+O(1/r^{D-2}).
\end{eqnarray}
The mass and tension computed from (\ref{charge}) 
agree with the expressions (\ref{2})
\footnote{Note that in computing  ${\mathcal T}$,
one should consider the integration over $\hat{S}^{z}=S^z/\Delta t$, 
with $\Delta t=\int dt$.
However, a similar problem appears also 
in the Hamiltonian approach \cite{Harmark:2004ch}.}.

One should  remark that the counterterm choice is not unique,
other choices being possible as well
(see \cite{Mann:2005cx} for a related discussion).
Our choice
of using (\ref{ct}) was motivated by the fact that the 
general expression for the
boundary stress-tensor is very simple. 
For an asymptotically ${\cal M}^{D-1}\times S^1$ spacetime,
we find that $I_{ct}$ (\ref{ct}) or other possible choices 
proposed in \cite{Kraus:1999di} regularizes
also
 the black strings Euclidean action,  yielding similar results to those obtained within the
background subtraction approach.

%This approach can  be extended to the KK black hole case.
%%%%%%%%%%%%%%%%%%%%%%%%%%%%%%%%%%%%%%%%%%%%%%%%%%%%
\subsection{Remarks on nonuniform bubble-like solutions }
%%%%%%%%%%%%%%%%%%%%%%%%%%%%%%%%%%%%%%%%%%%%%%%%%%%%
There is a also a simple way to generate time dependent
solutions with a nontrivial extradimension dependence.
A Lorentzian solution of the vacuum Einstein equations
is found
by using the following analytic continuation in the general configuration (\ref{metric})
(with $d\Omega_{D-3}^2=d\theta^2+\sin^2\theta d\Omega_{D-4}^2$)
\begin{eqnarray}
t\equiv i \chi,~~\theta-\pi/2\equiv i\tau.
\label{ac}
\end{eqnarray}
The new solution reads
\begin{eqnarray} 
\label{metricb}
ds^2=e^{2A(r,z)}f(r)d\chi^2+e^{2B(r,z)}\left(\frac{dr^2}{f(r)}+dz^2\right)+e^{2C(r,z)}r^2
(-d\tau^2+\cosh^2 \tau d\Omega_{D-4}^2),
\end{eqnarray}
with $A(r,z),~B(r,z),~C(r,z)$ and $f(r)$ the functions of the black string configuration.

This technique of double analytic continuation was originally developed
for the study of the stability of the KK vacuum \cite{Witten:1981gj},
 \cite{Dowker:1995gb},
and has been considered in the last years by many authors in AdS/CFT context
(see e.g. \cite{Birmingham:2002st}).

However, the solution one finds starting with a nonuniform vacuum black string  
has some new features, 
as compared to the Schwarzschild-Tangherlini seed solution.
In that case, the resulting spacetime 
describes a contracting and then expanding "bubble of nothing".
The new configuration (\ref{metricb}) has now two compact extradimensions $z$ and $\chi$.
The radial variable $r$ is restricted
to the range $r \geq r_0$.
$r=r_0$ is not the boundary of spacetime, but it is the $S^{D-4}\times S^1$ of minimal area.
The curves at
$r=r_0$ with constant $z$ (and constant points on $S^{D-4}$) are geodesics.
Also, regularity
at $r=r_0$ requires $\chi$ to be periodic with period
$\beta=1/T$, with $T$ given by (\ref{temp}).
One can see that the geometry traced out by the $r=r_0$ surface 
is the $(D-3)$-dimensional de Sitter spacetime with a $z-$dependent
conformal factor, times the extradimension.
Thus (\ref{metricb}) would rather describe a nonuniform "vortex of nothing".

One can use the same techniques as in Section 4 to find the corresponding 
$D-$dimensional solutions in Einstein-Maxwell-dilaton theory.

{
KK bubble solutions in $d=5,~6$ dimensions
have been also considered in Ref. \cite{Elvang:2004iz}. 
A number of exact solutions have been presented there, describing
sequences of KK bubbles and black holes, placed alternately
so that the black holes are held apart by the bubbles.
However,  
the configurations in \cite{Elvang:2004iz} differ from
(\ref{metricb}),
since they are static and asymptotically approach
Minkowski spacetime times a circle.
}

%%%%%%%%%%%%%%%%%%%%%%%%%%%%%%%%%%%%%%%%%%%%%%%%%
\section{Numerical nonuniform black string solutions}
%%%%%%%%%%%%%%%%%%%%%%%%%%%%%%%%%%%%%%%%%%%%%%%%%  

%%%%%%%%%%%%%%%%%%%%%%%%%%%%%%%%%%%%%%%%%%%%%%%%%%%%
\subsection{Numerical procedure}
%%%%%%%%%%%%%%%%%%%%%%%%%%%%%%%%%%%%%%%%%%%%%%%%%%%%

Our main concern here is the numerical construction of 
nonuniform black string solutions in $D=5$ dimensions.
We have also constructed nonuniform black string solutions
in $D=6$ dimensions,
reproducing and extending the known set of solutions,
obtained previously by different methods, 
which could not be successfully applied in $D=5$ dimensions
\cite{Wiseman:2002zc,Wiseman:2002ti,Kol:2003ja,Kudoh:2004hs}.

To obtain nonuniform black string solutions,
we solve the set of three coupled non-linear
elliptic partial differential equations numerically \cite{schoen},
subject to the above boundary conditions.
We employ dimensionless coordinates
$\bar r$ and $\bar z$,
\begin{equation}
\bar r = \tilde r/(1+\tilde r)  , \ \ \ \bar z = z/ L  ,
  \label{barx2} \end{equation}
where the compactified radial coordinate $\bar r$
maps spatial infinity to the finite value $\bar r=1$,
and $L$ is the asymptotic length of the compact direction.
The numerical calculations are based on the Newton-Raphson method
and are performed with help of the program FIDISOL \cite{schoen},
which provides also an error estimate for each unknown function.

The equations are discretized on a non-equidistant grid in
$\bar r$ and $\bar z$.
Typical grids used have sizes $65 \times 50$,
covering the integration region
$0\leq \bar r \leq 1$ and $0\leq \bar z \leq 1/2$.
(See \cite{schoen} and \cite{kk}
for further details and examples for the numerical procedure.)
For the nonuniform strings the estimated relative errors 
range from approximately $\approx 0.001$\%
for small geometric deformation
to $\approx 1$\% for large deformation.
Further discussion of the numerical accuracy is deferred to Appendix A.

The horizon coordinate $r_0$ and the asymptotic length $L$ of the compact
direction enter the equations of motion as parameters.
The results presented are mainly obtained with the parameter choice
\begin{equation}
r_0=1 \ , \ \ \ L=L^{\rm crit}= \left\{ \begin{array}{ccc}
7.1713 & & D=5\\
4.9516 & & D=6
\end{array} \right.
, \label{r_0-L} \end{equation}
where $L^{\rm crit}$ represents the value, where the instability of the uniform
string occurs.
The branch of nonuniform strings is then obtained by
starting at the critical point of the uniform string branch
and varying the boundary parameter $d_0$, which enters the Eq.~(\ref{bc2}),
relating the values of the functions $A$ and $B$ at the horizon.

%%%%%%%%%%%%%%%%%%%%%%%%%%%%%%%%%%%%%%%%%%%%%%%%%%%%
\subsection{Black string properties}
%%%%%%%%%%%%%%%%%%%%%%%%%%%%%%%%%%%%%%%%%%%%%%%%%%%%

Let us first consider the metric functions $A$, $B$ and $C$
for nonuniform string solutions
as functions of the radial coordinate $r$
and of the coordinate $z$ of the compact direction.
Keeping the asymptotic length $L$ of the compact direction
and the horizon coordinate $r_0$ fixed, 
the solutions change smoothly with boundary parameter $d_0$.
A measure of the deformation of the solutions is given by the
nonuniformity parameter $\lambda$ \cite{Gubser:2001ac}
\begin{equation} 
\lambda = \frac{1}{2} \left( \frac{{\cal R}_{\rm max}}{{\cal R}_{\rm min}}
 -1 \right)
, \label{lambda} \end{equation}
where ${\cal R}_{\rm max}$ and ${\cal R}_{\rm min}$
represent the maximum radius of a $(D-3)$-sphere on the horizon and
the minimum radius, being the radius of the `waist'.
Thus for uniform black strings $\lambda=0$,
while the conjectured horizon topology
changing transition should be approached
for $\lambda \rightarrow \infty$ 
\cite{Kol:2003ja,Wiseman:2002ti}.
As $d_0$ first increases and then decreases again,
the nonuniformity parameter $\lambda$ 
increases monotonically.

In Figure 1 we exhibit the metric functions for
$D=5$ nonuniform string solutions for
several values of the nonuniformity parameter,
$\lambda=1,~2,~5,~9$.
The functions exhibit extrema on the symmetry axis $z=0$
at the horizon. 
As $\lambda$ increases, the extrema increase in height and become
increasingly sharp (implying a deteriorating numerical accuracy
for large values of $\lambda$).

To obtain a more quantitative picture of the metric functions,
we exhibit $-g_{tt}/f=e^{2A}$, $g_{zz}=e^{2B}$, and
$g_{\theta \theta}/r^2=e^{2C}$ in Figure 2 for several fixed values
of $z$.

The spatial embedding of the horizon into 3-dimensional space
is shown in Figure 3 for the $D=5$ nonuniform black string solutions.
In these embeddings the proper radius of the horizon is plotted
against the proper length along the compact direction,
yielding a geometrical view of the nonuniformity
of the solutions.

The deformation of the horizon of the $D=5$ nonuniform black string solutions
is further explored in Figure 4. 
The maximum radius of the 2-sphere on the horizon ${\cal R}_{\rm max}$ 
and the
%%%%%%%%%%%%%%%%%%%%%%%%%%%%%%%%%%%%%%%%%%%%%%%%%%%%%%%%%%%%%%%%

\setlength{\unitlength}{1cm}
\begin{picture}(15,18)
\put(-1,0){\epsfig{file=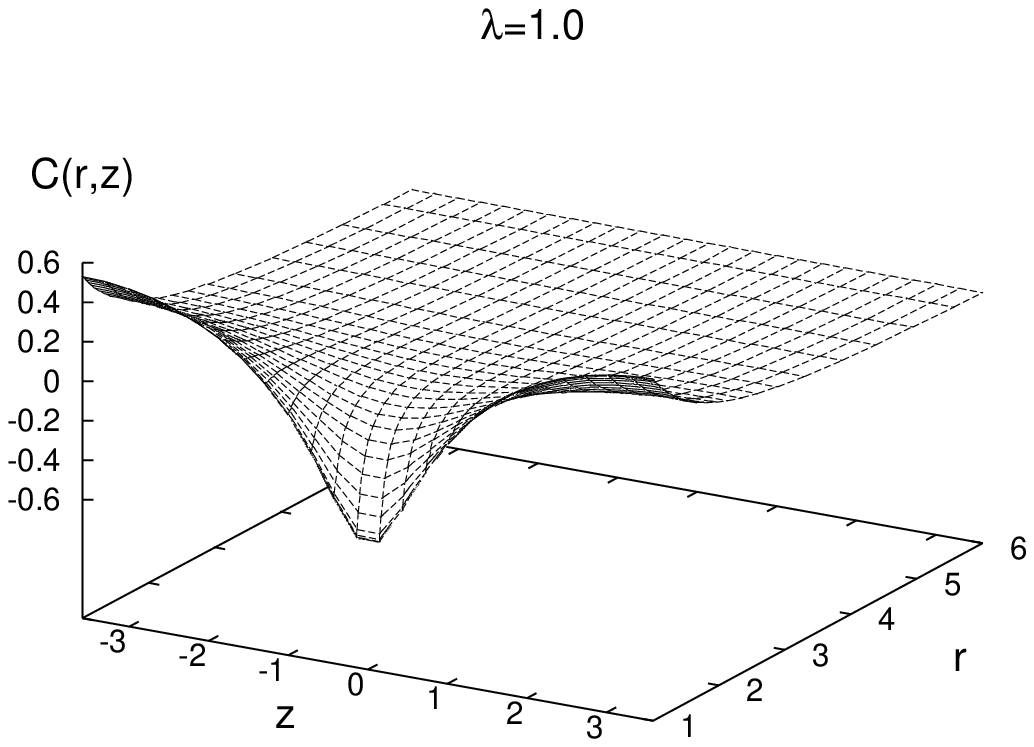,width=8cm}}
\put(7,0){\epsfig{file=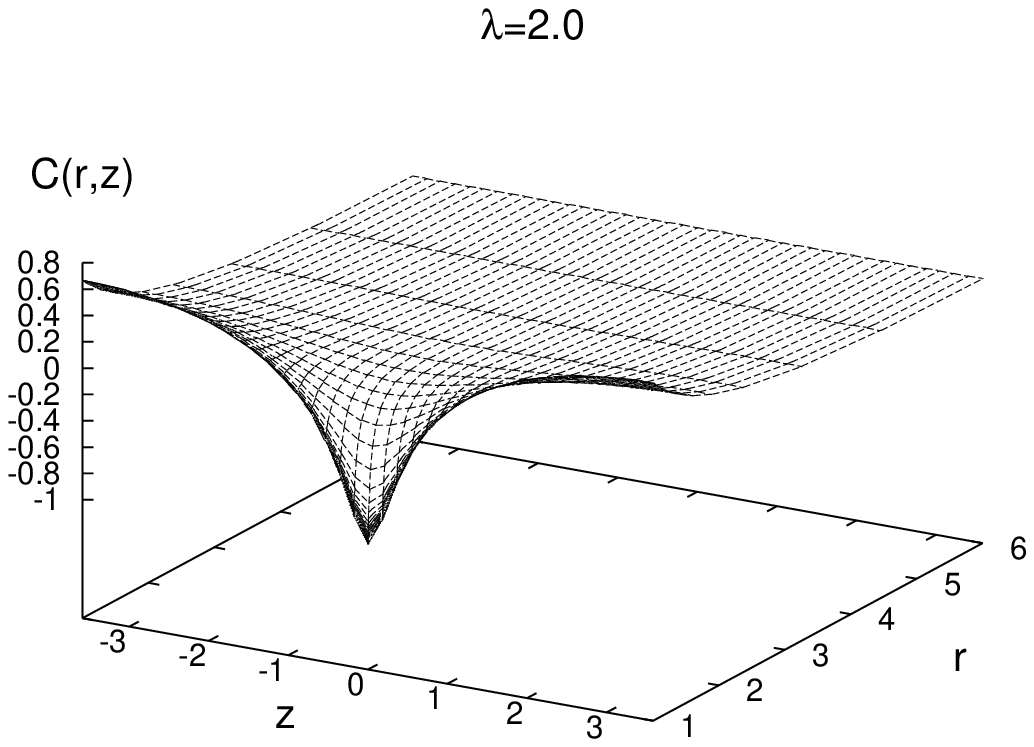,width=8cm}}
\put(-1,6){\epsfig{file=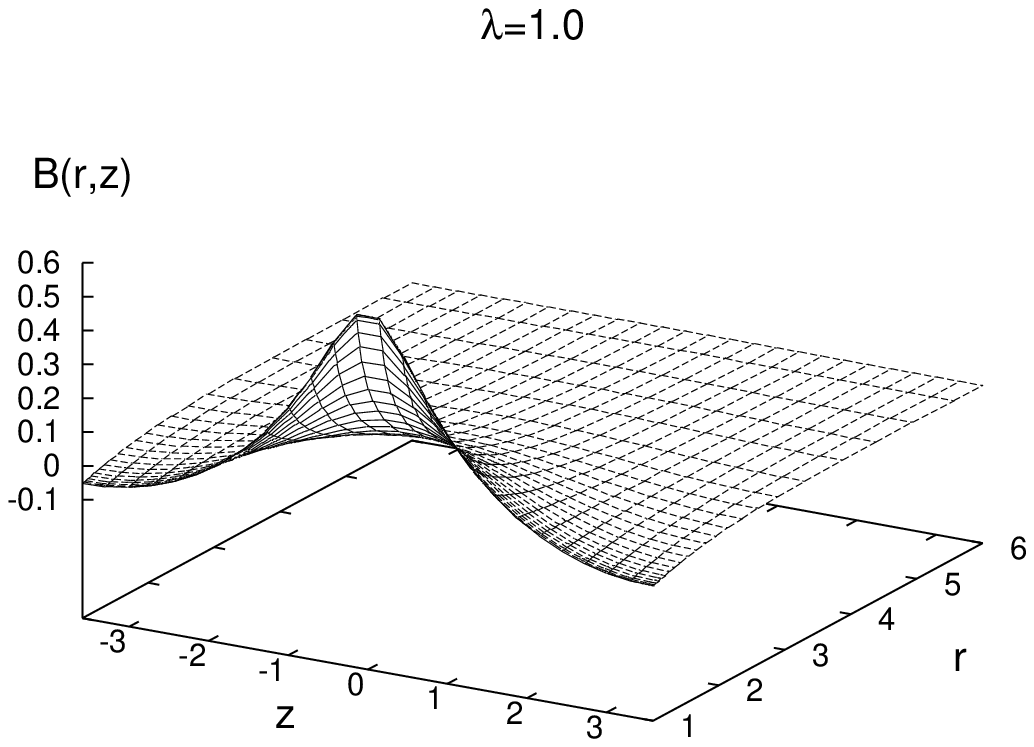,width=8cm}}
\put(7,6){\epsfig{file=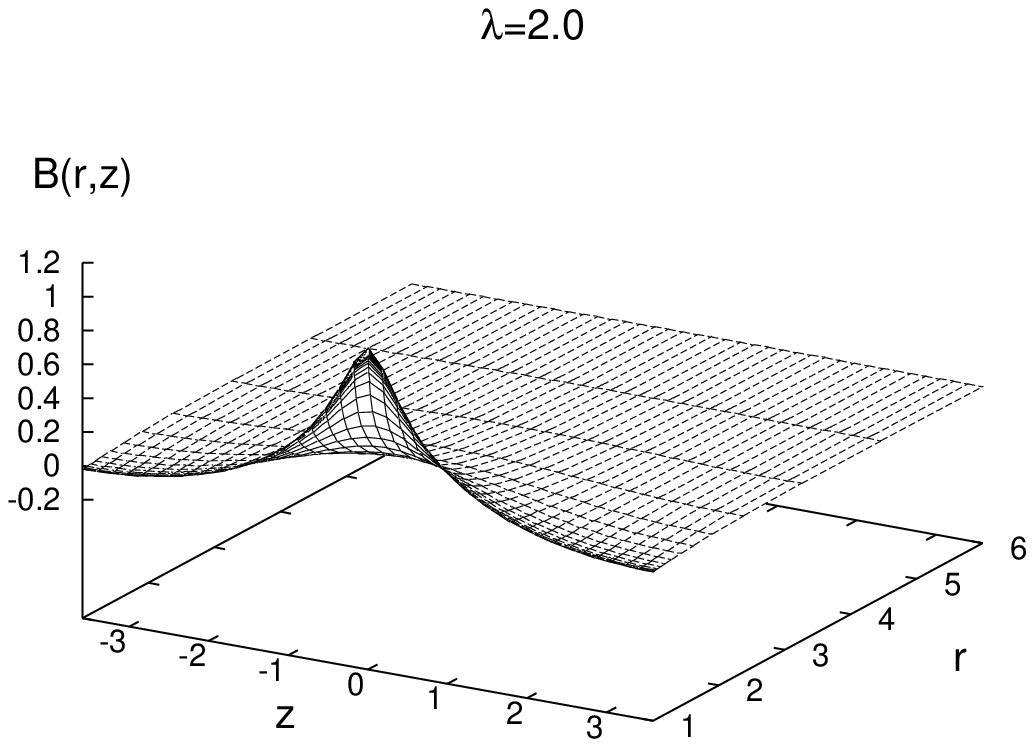,width=8cm}}
\put(-1,12){\epsfig{file=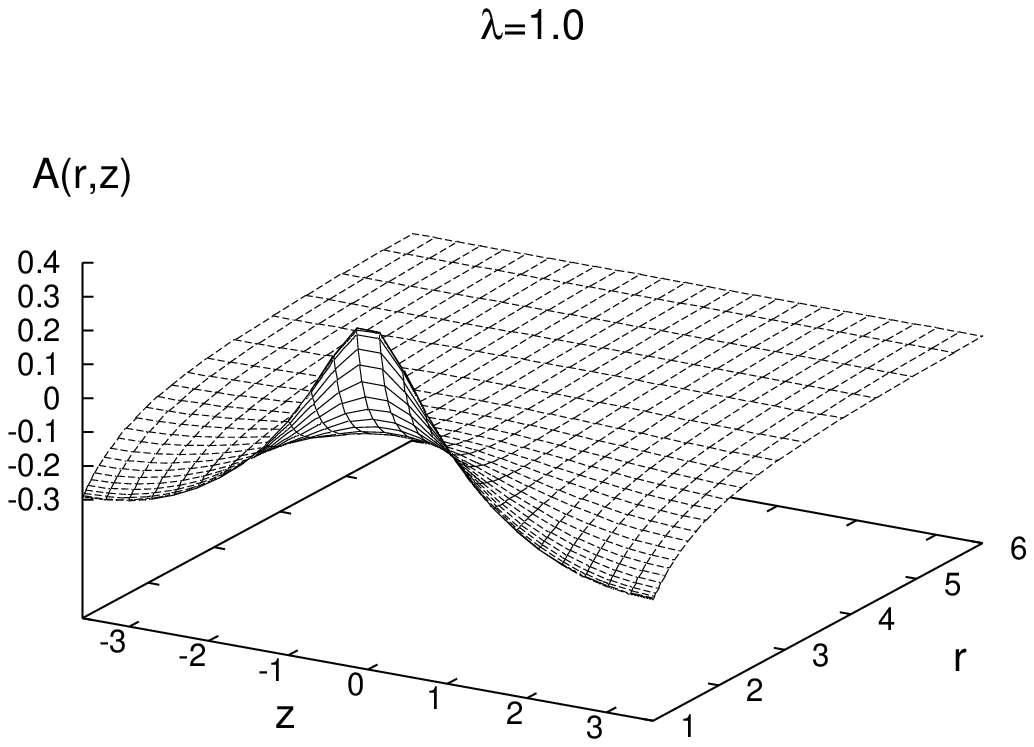,width=8cm}}
\put(7,12){\epsfig{file=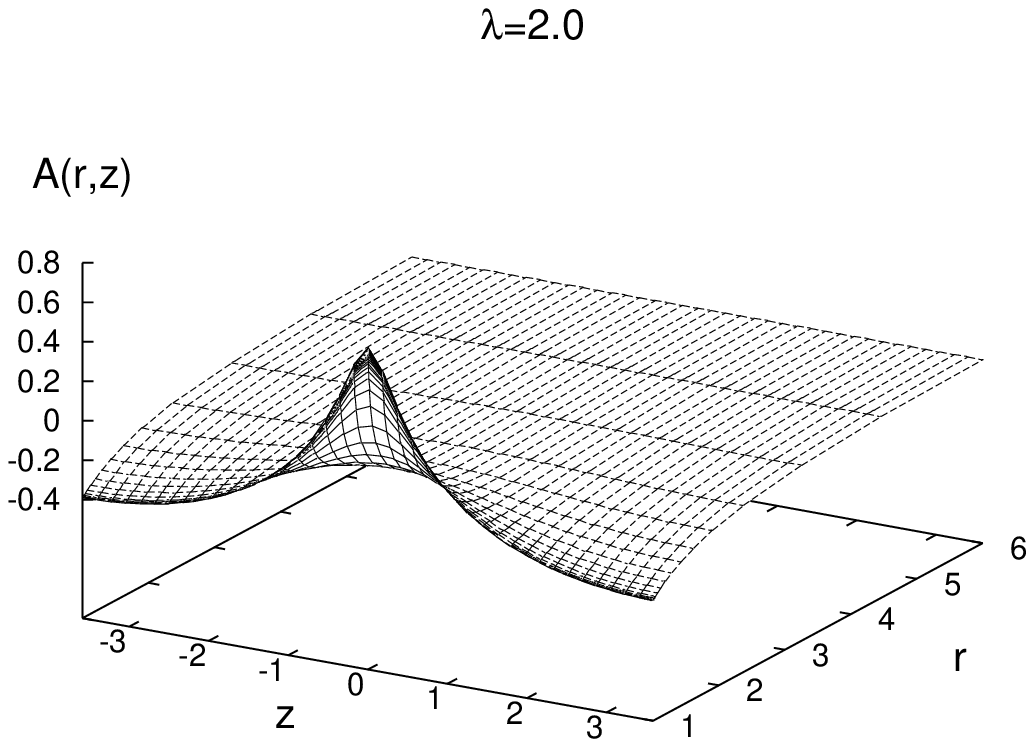,width=8cm}}
\end{picture}
\\
\\
{\small {\bf Figure 1.}
The metric functions $A$, $B$ and $C$ 
of the $D=5$ nonuniform string solutions
are shown as functions of the radial coordinate $r$,
with the horizon located at $r_0 =1$,
and the coordinate $z$ of the compact direction
with asymptotic length $L=L^{\rm crit}=7.1713$,
for several values of the nonuniformity parameter,
$\lambda=1$ (first column),
$\lambda=2$ (second column),
$\lambda=5$ (third column),
$\lambda=9$ (forth column).
}
\vspace{0.5cm}
\\
\\
%%%%%%%%%%%%%%%%%%%%%%%%%%%%%%%%%%%%%%%%%%%%%%%%%%%
\setlength{\unitlength}{1.cm}
\begin{picture}(15,18)
\put(-1,0){\epsfig{file=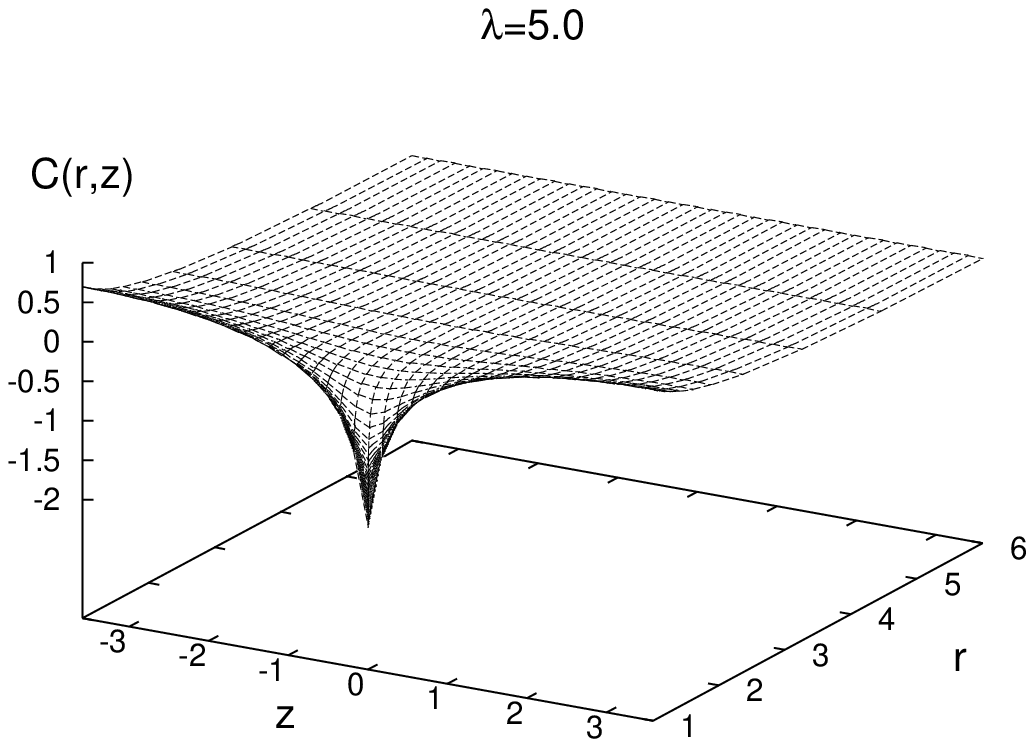,width=8.2cm}}
\put(7,0){\epsfig{file=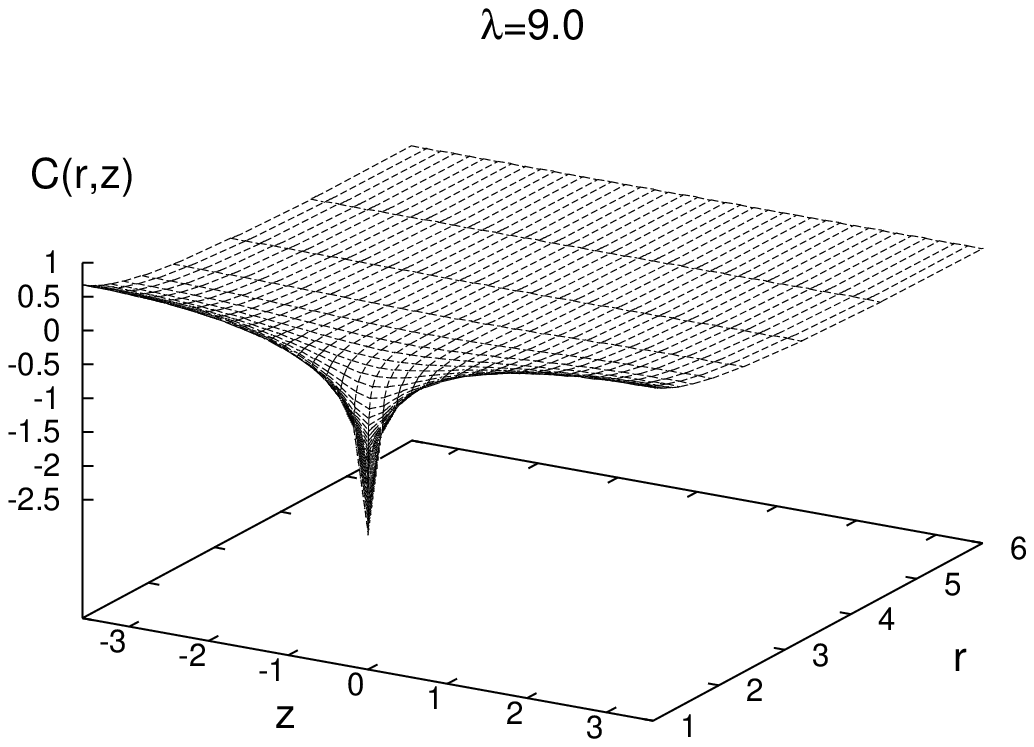,width=8.2cm}}
\put(-1,6){\epsfig{file=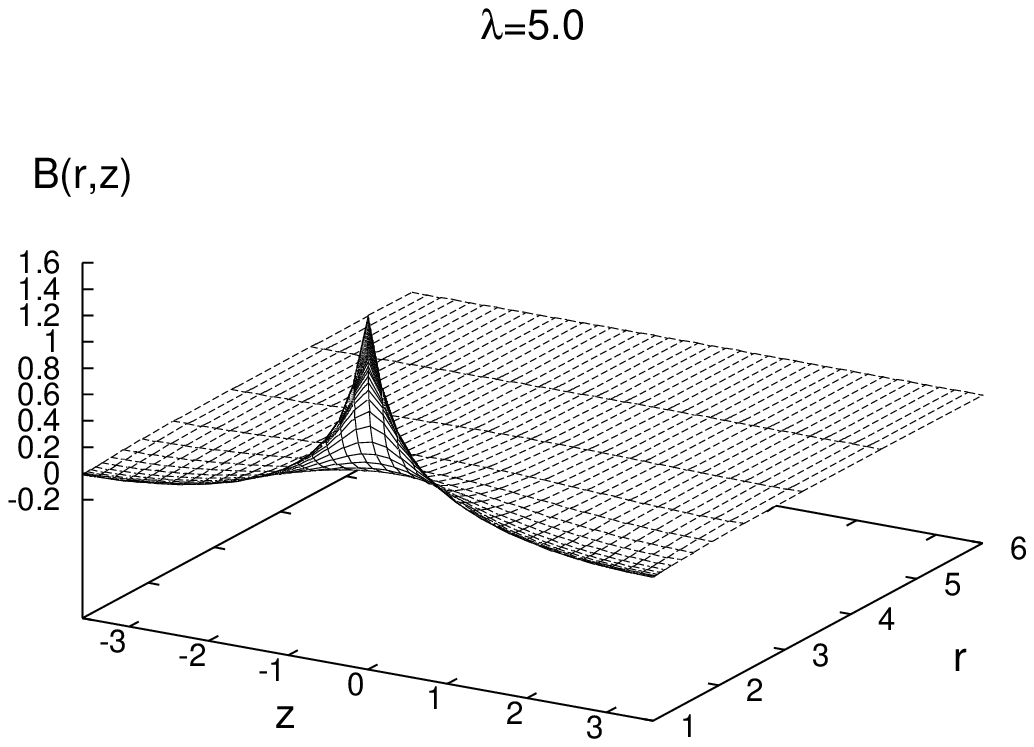,width=8.2cm}}
\put(7,6){\epsfig{file=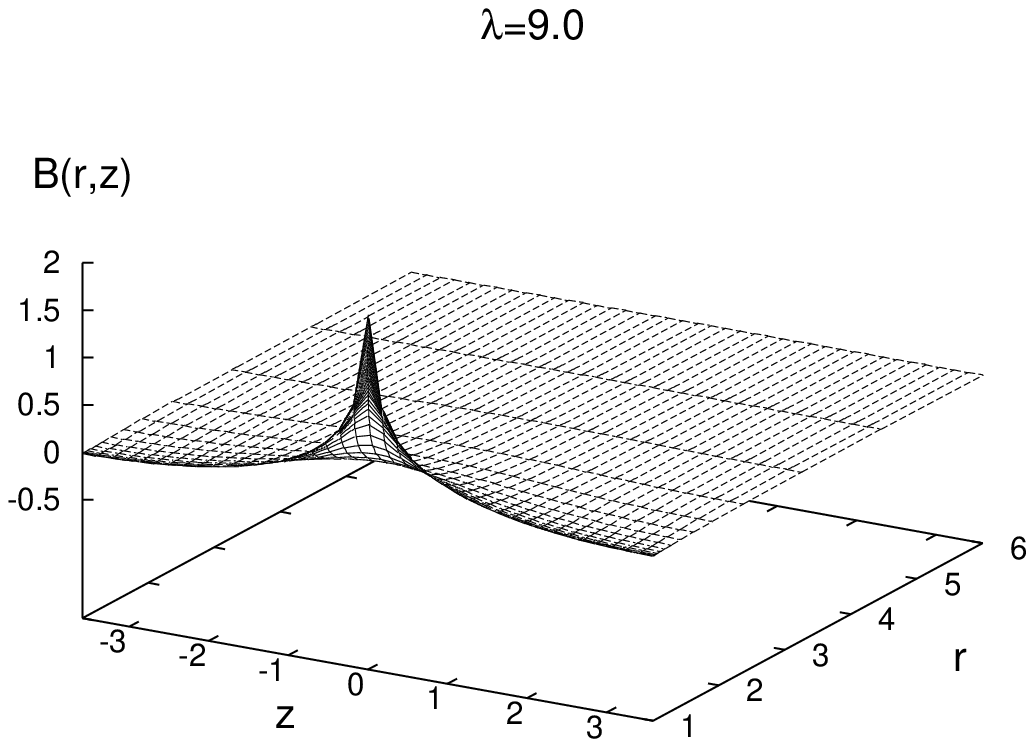,width=8.2cm}}
\put(-1,12){\epsfig{file=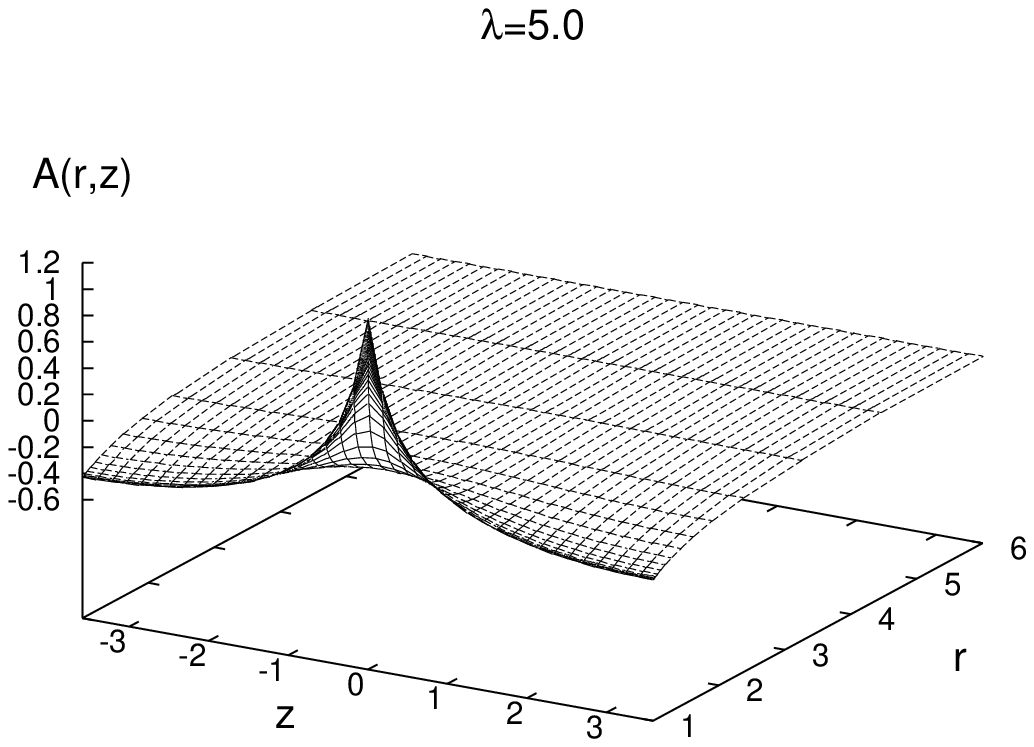,width=8.2cm}}
\put(7,12){\epsfig{file=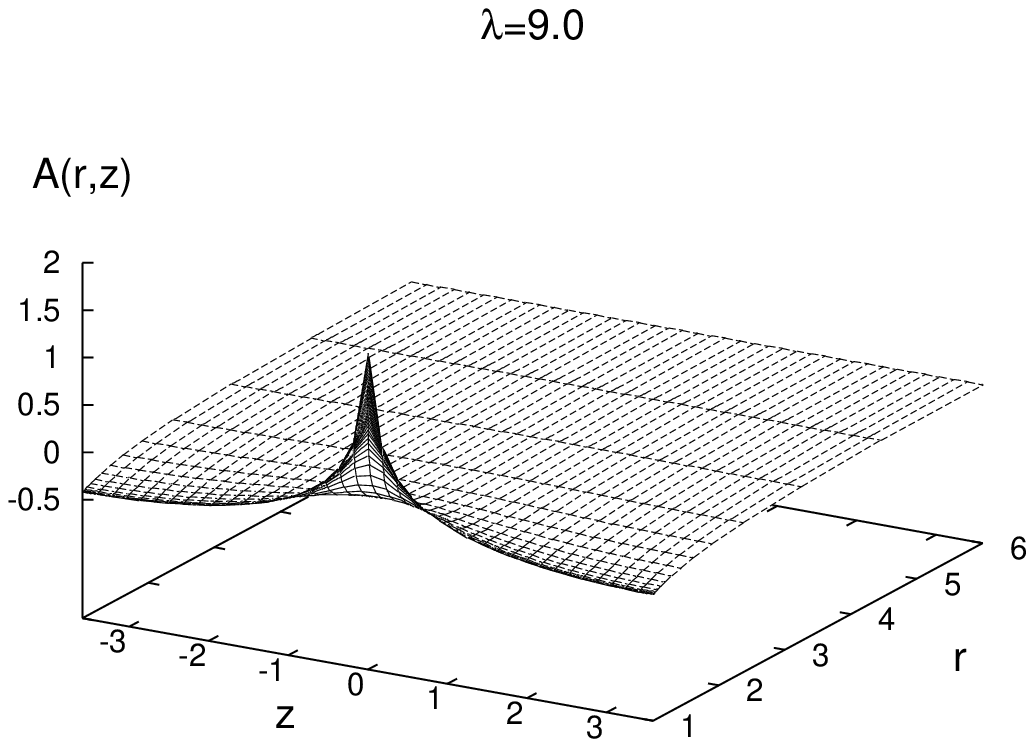,width=8.2cm}}
\end{picture}
\\
\\
\\
{\small {\bf Figure 1.}
Figure 1 continued.
}
\vspace{0.5cm}
\\

\setlength{\unitlength}{1cm}
\begin{picture}(15,18)
\put(-1,0){\epsfig{file=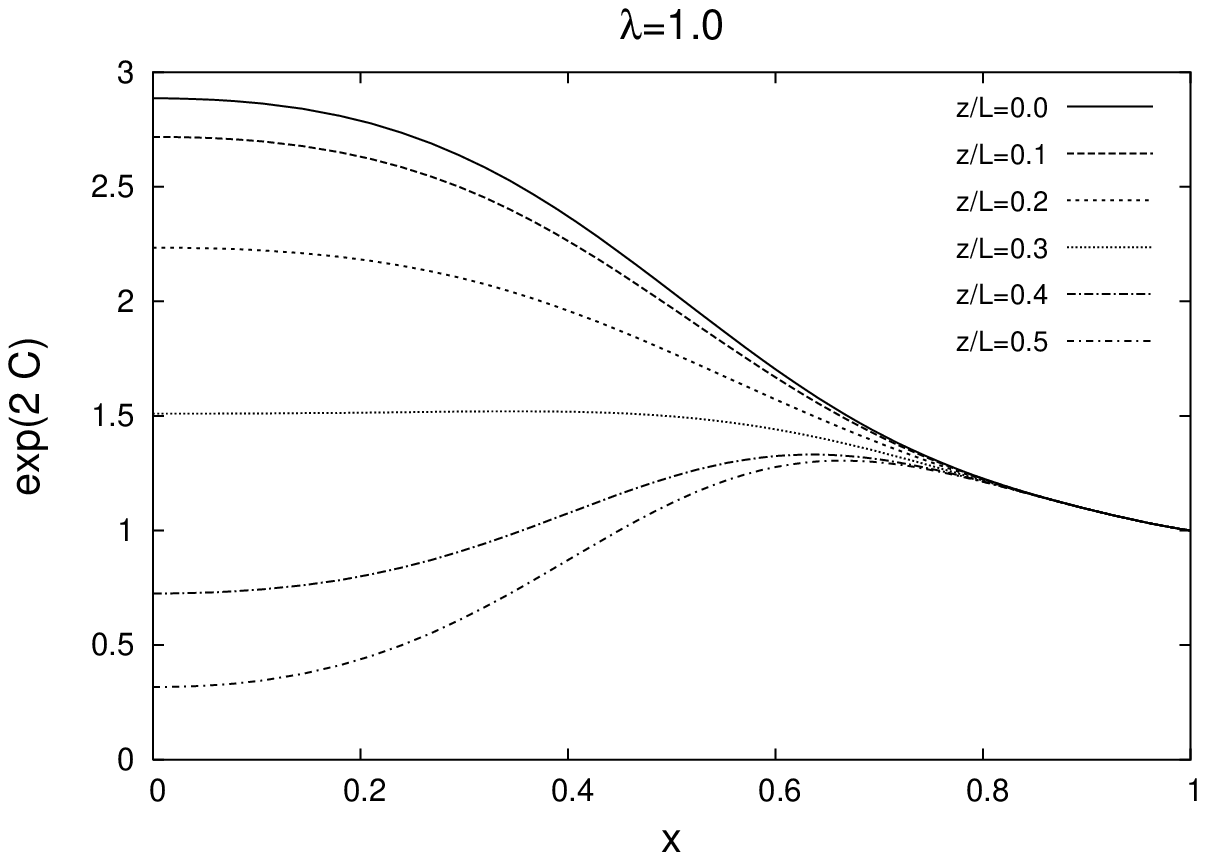,width=8cm}}
\put(7,0){\epsfig{file=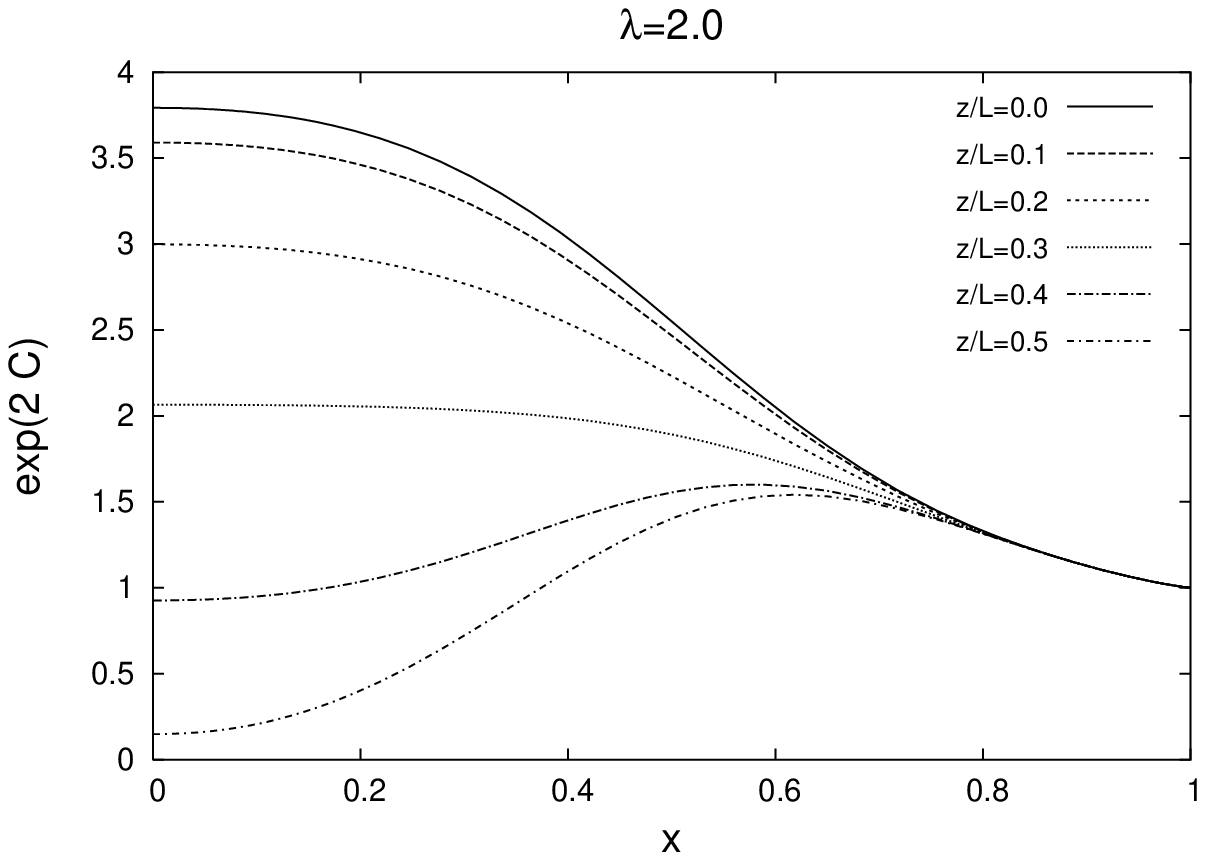,width=8cm}}
\put(-1,6){\epsfig{file=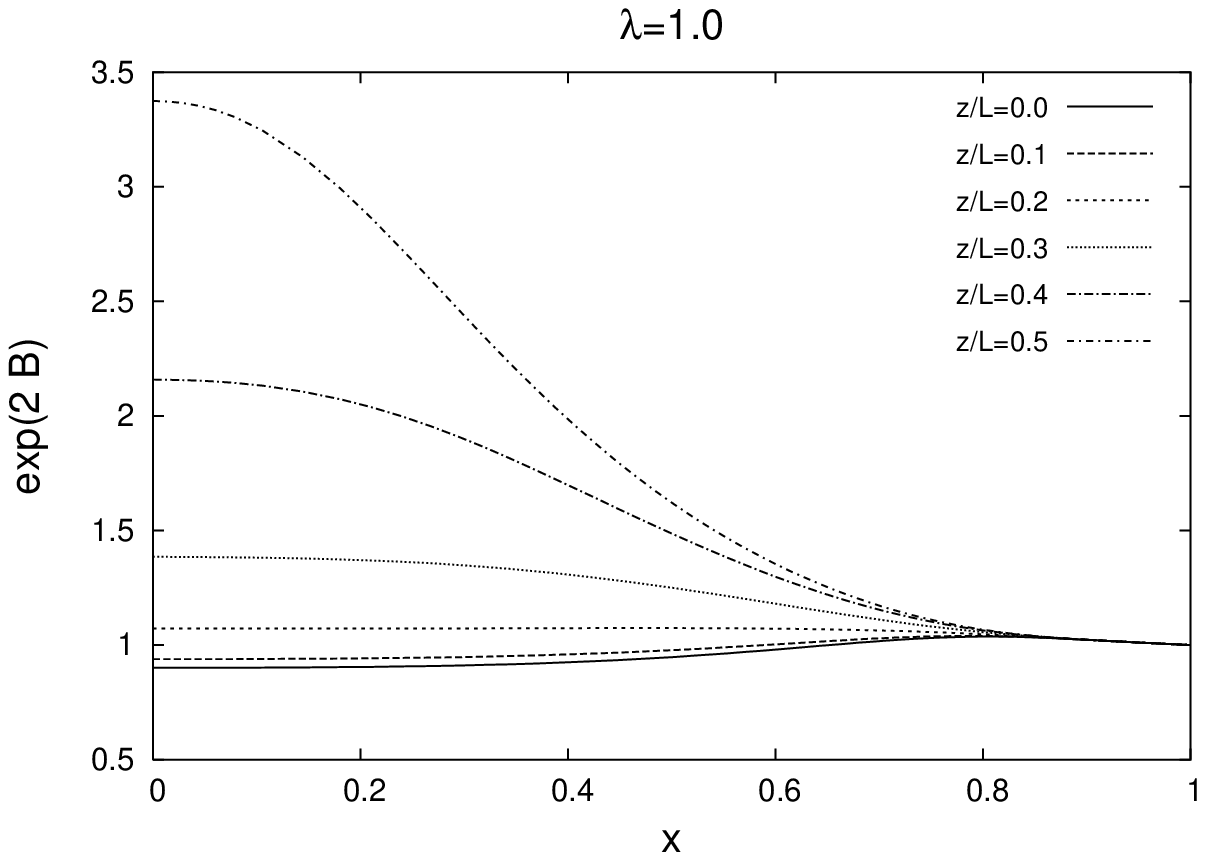,width=8cm}}
\put(7,6){\epsfig{file=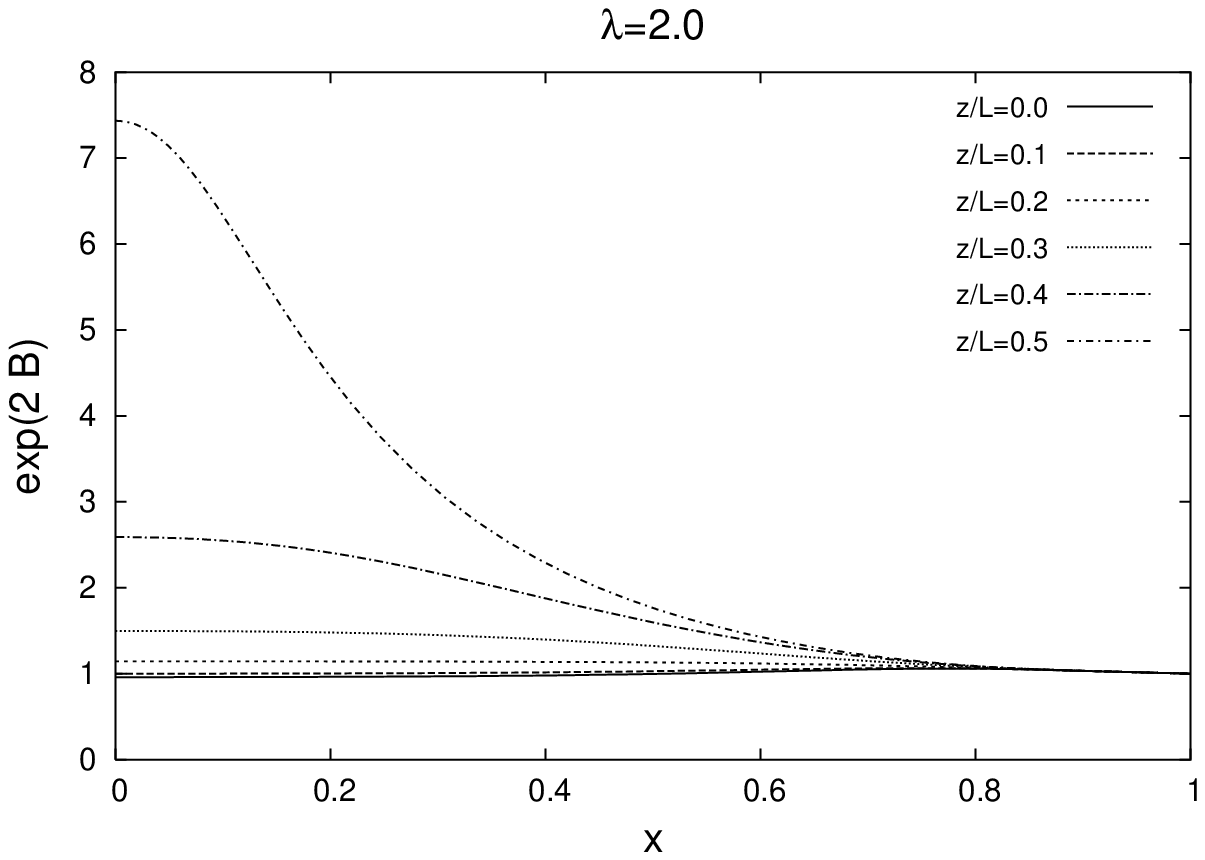,width=8cm}}
\put(-1,12){\epsfig{file=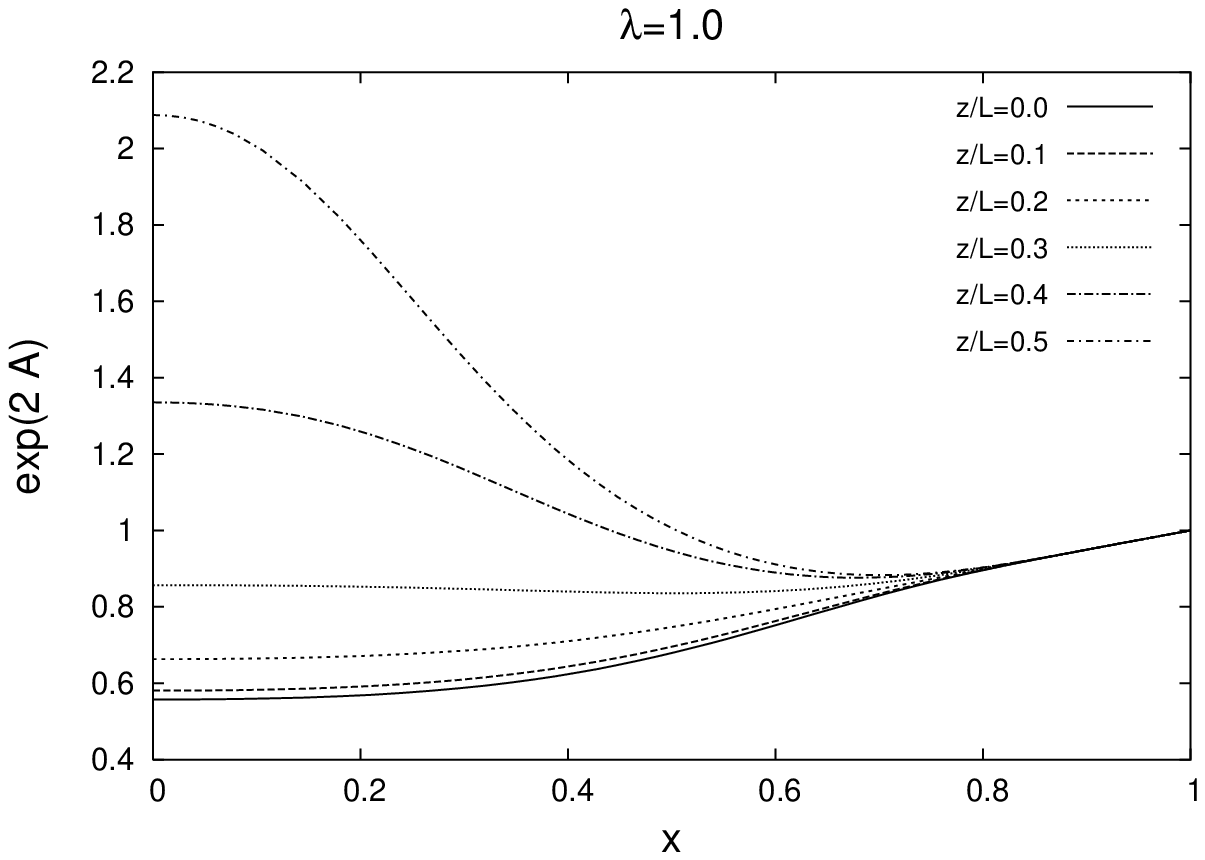,width=8cm}}
\put(7,12){\epsfig{file=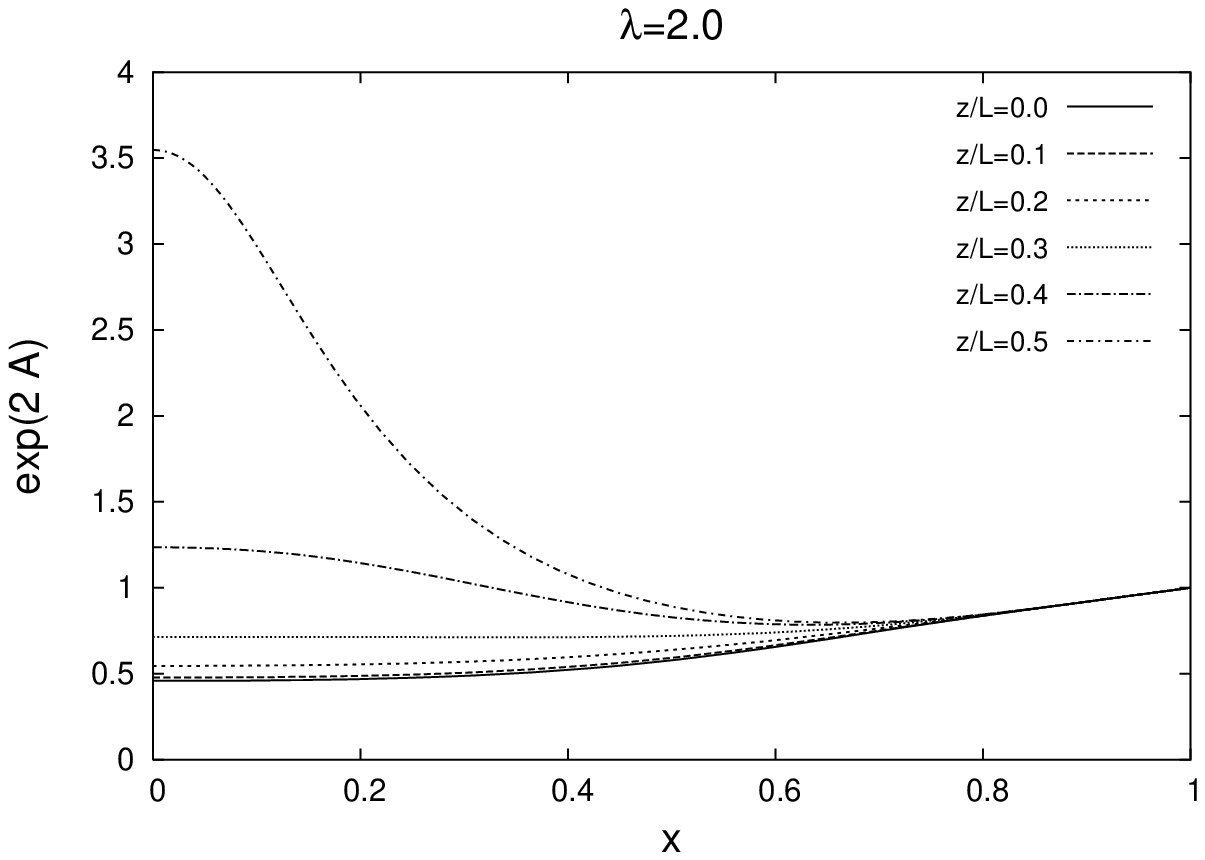,width=8cm}}
\end{picture}
\\
\\
{\small {\bf Figure 2.}
The metric functions 
$e^{2A}$, $e^{2B}$, and $e^{2C}$
of the $D=5$ nonuniform string solutions
are shown as functions of the compactified radial coordinate $x= \bar r$,
for several fixed values of the coordinate $z$ of the compact direction
($z/L=0,~0.1,~0.2,~0.3,~0.4,~0.5$),
as well as of the nonuniformity parameter $\lambda$
($\lambda=1$ (first column),
$\lambda=2$ (second column),
$\lambda=5$ (third column),
$\lambda=9$ (fourth column)).
}
\vspace{0.5cm}
\\
\setlength{\unitlength}{1cm}
\begin{picture}(15,18)
\put(-1,0){\epsfig{file=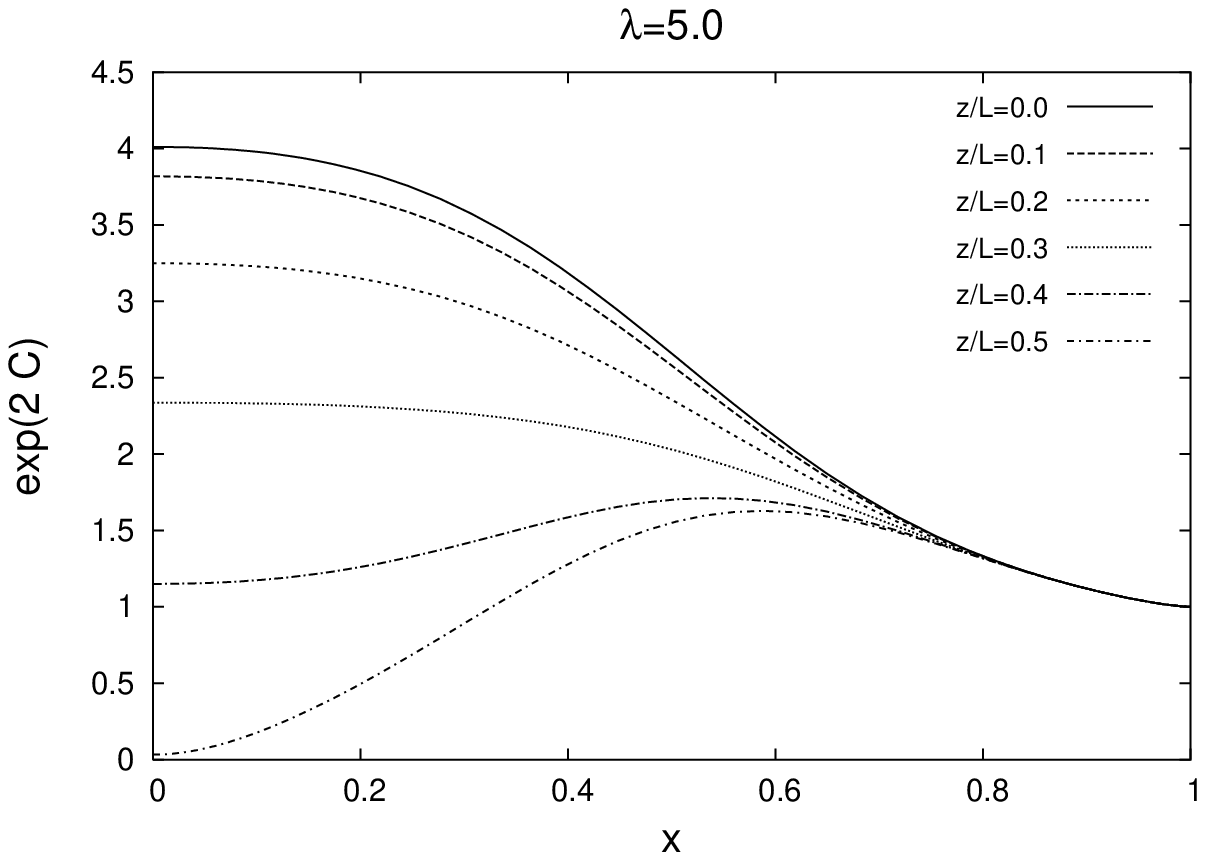,width=8cm}}
\put(7,0){\epsfig{file=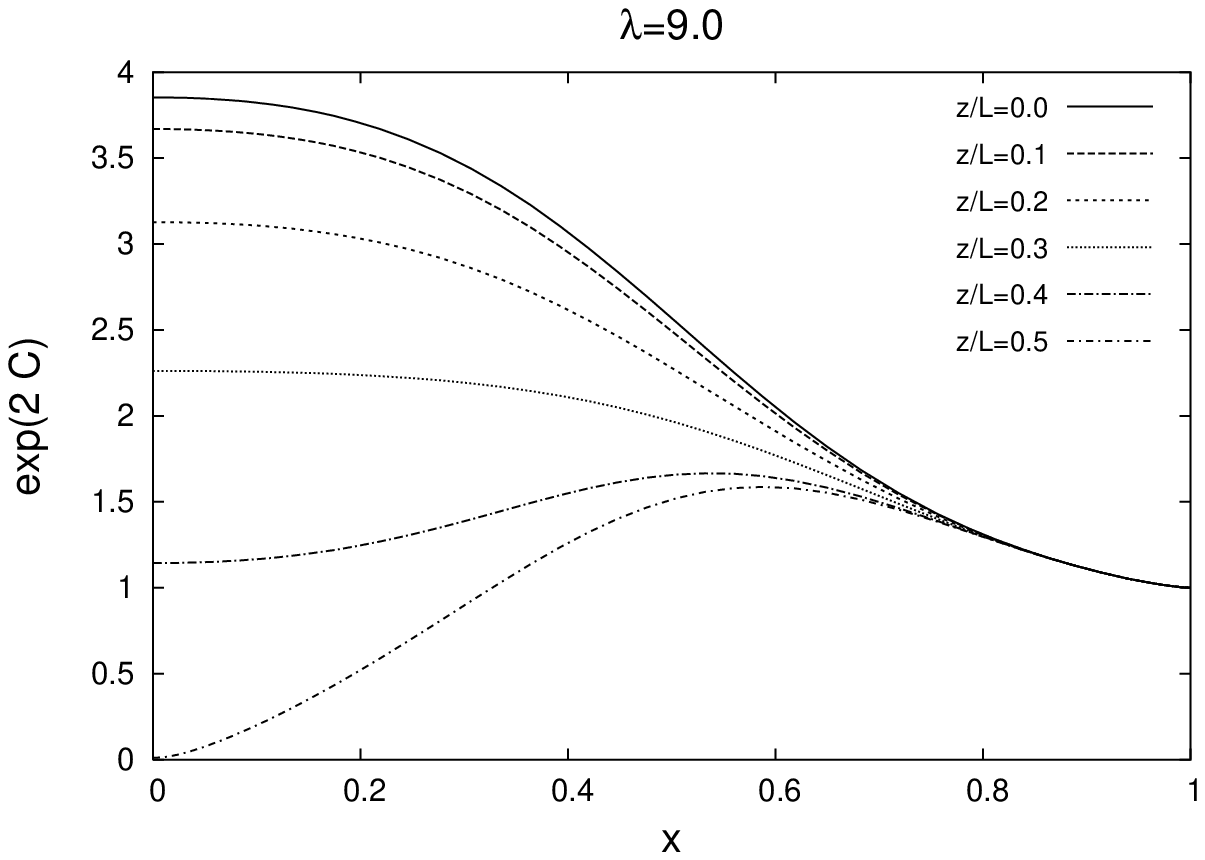,width=8cm}}
\put(-1,6){\epsfig{file=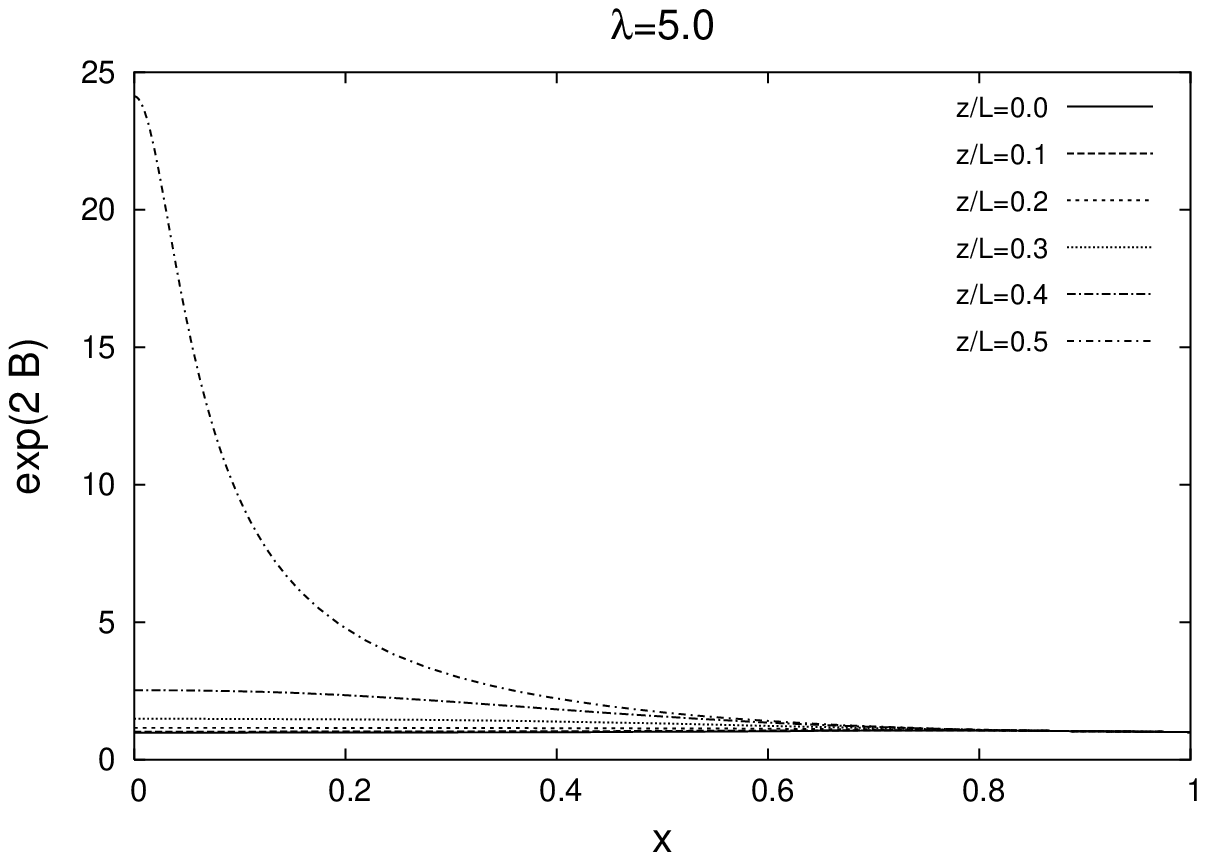,width=8cm}}
\put(7,6){\epsfig{file=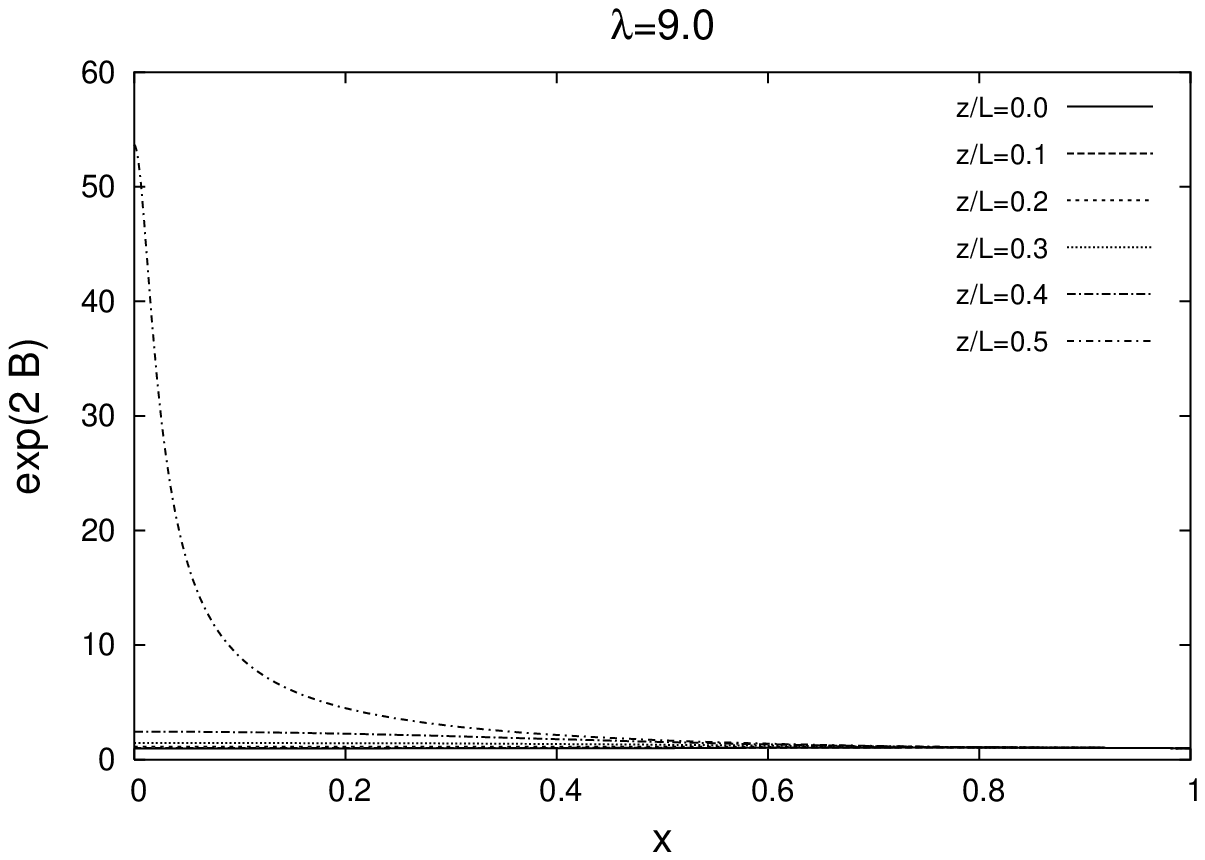,width=8cm}}
\put(-1,12){\epsfig{file=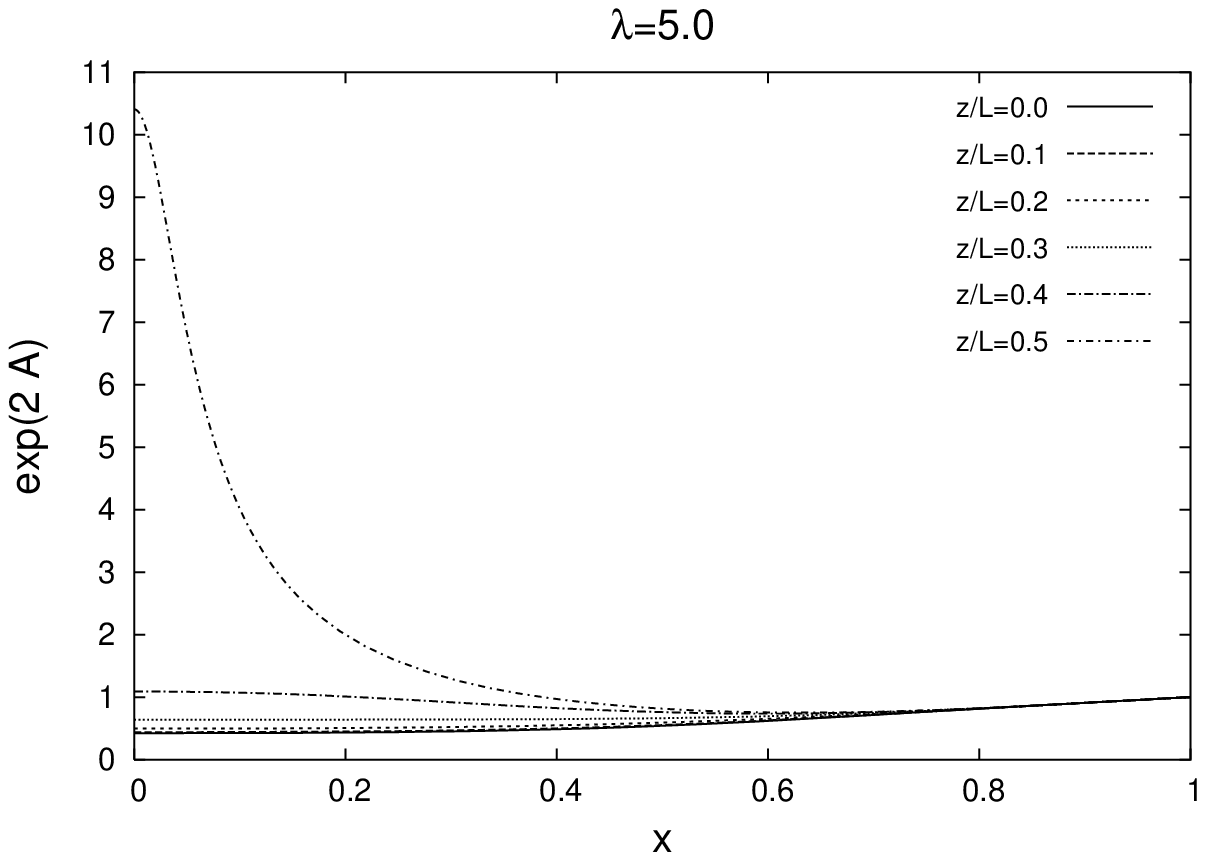,width=8cm}}
\put(7,12){\epsfig{file=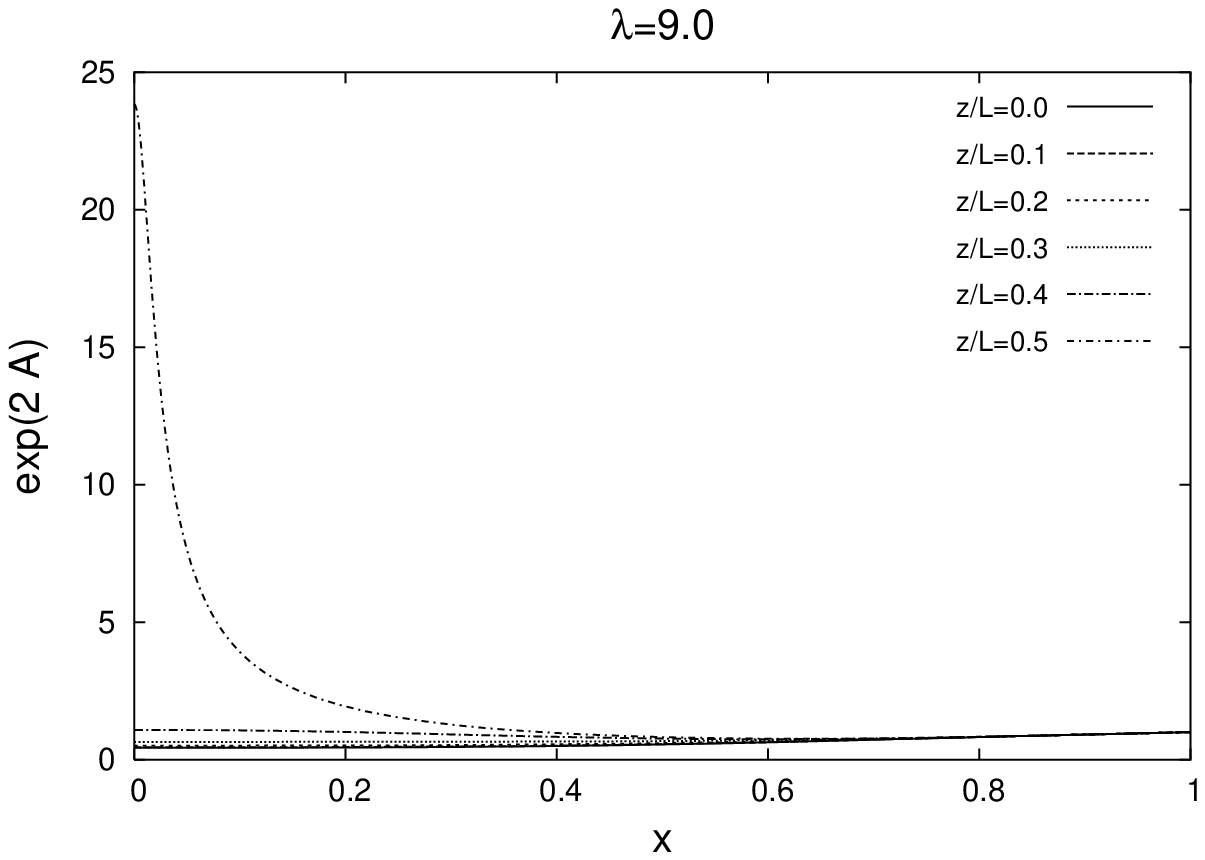,width=8cm}}
\end{picture}
\\
\\
\\
{\small {\bf Figure 2.}
Figure 2 continued.
}
\vspace{0.5cm}
\\

\setlength{\unitlength}{1cm}
\begin{picture}(15,18)
\put(-1,0){\epsfig{file=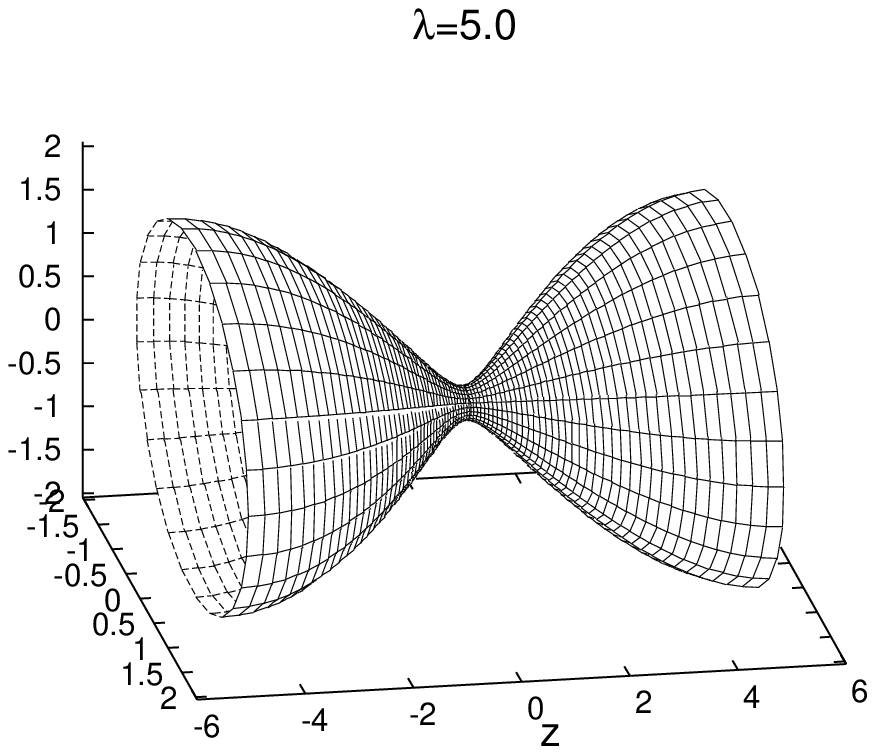,width=8cm}}
\put(7,0){\epsfig{file=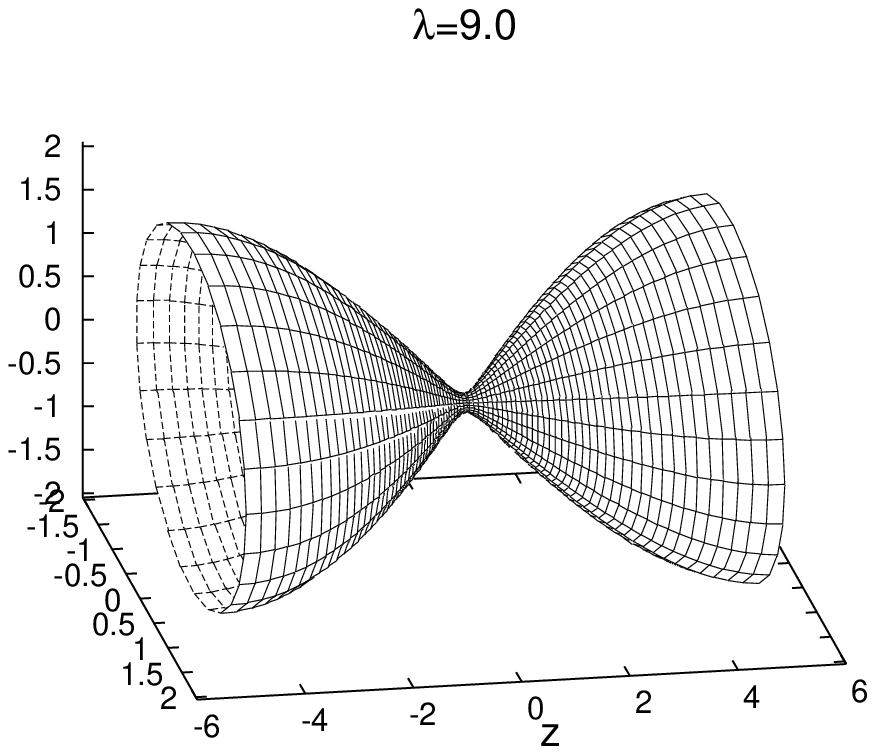,width=8cm}}
\put(-1,6){\epsfig{file=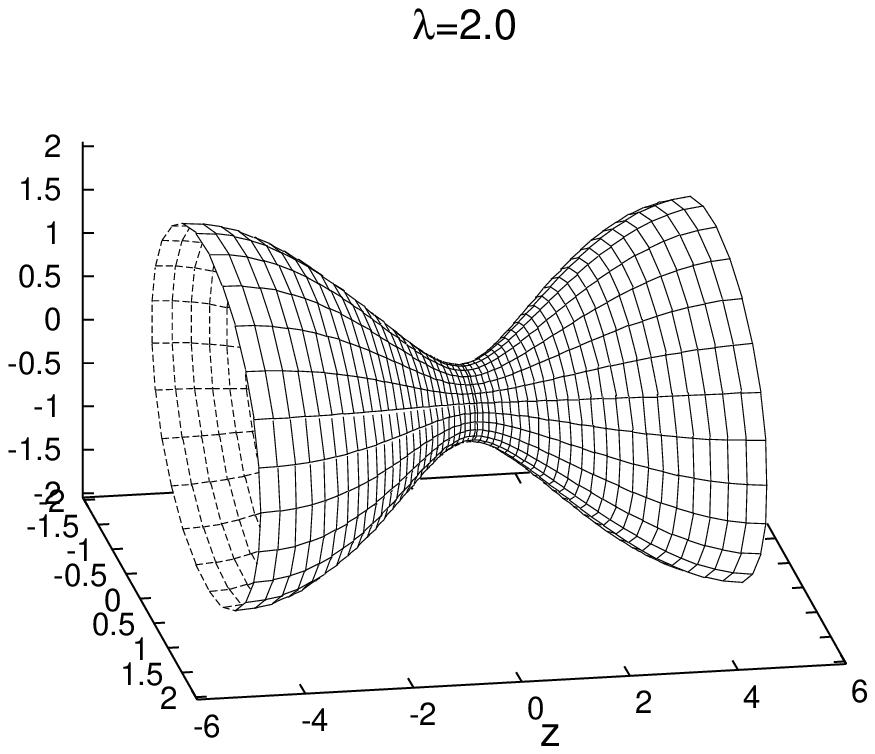,width=8cm}}
\put(7,6){\epsfig{file=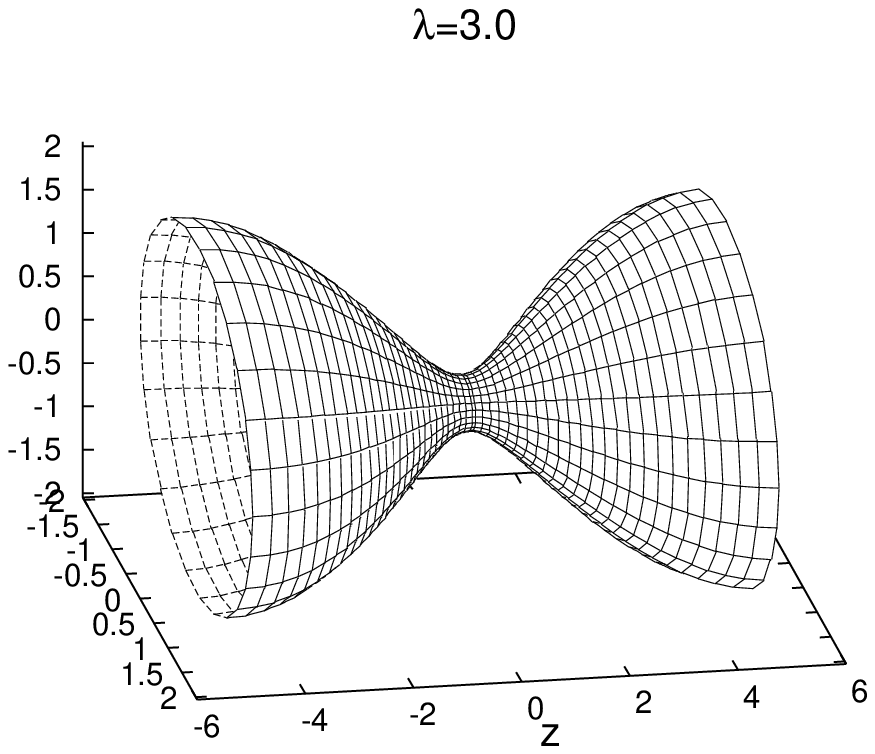,width=8cm}}
\put(-1,12){\epsfig{file=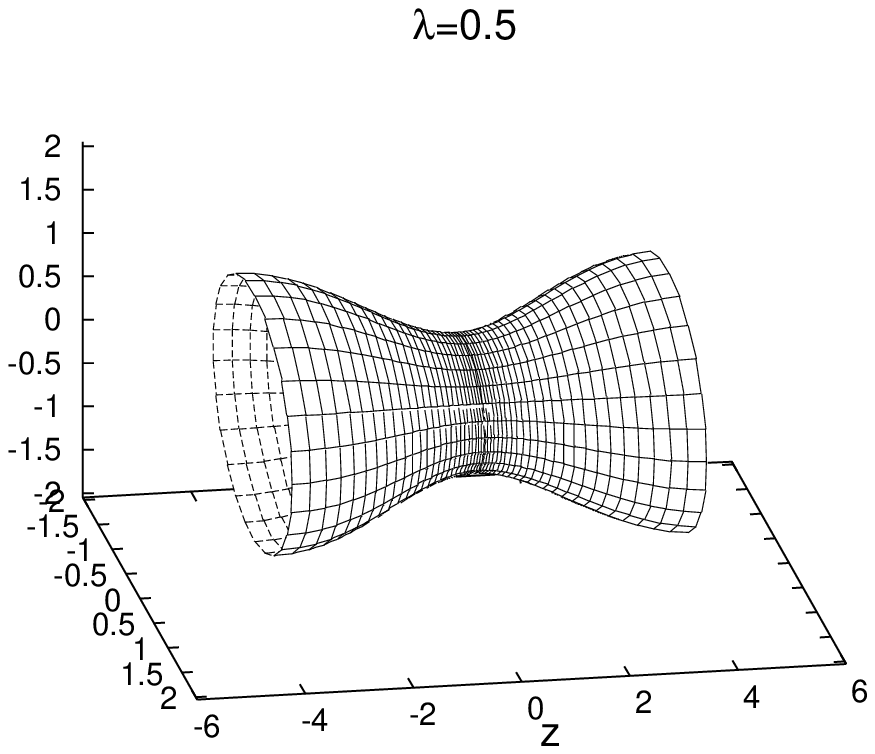,width=8cm}}
\put(7,12){\epsfig{file=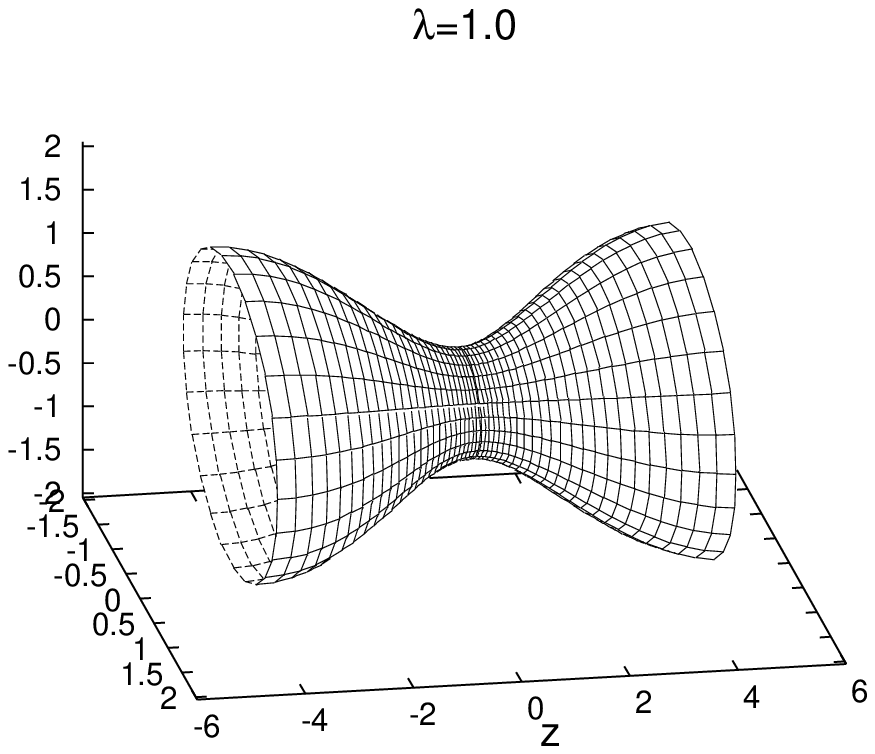,width=8cm}}
\end{picture}
\\
\\
\\
{\small {\bf Figure 3.}
The spatial embedding of the horizon
of $D=5$ nonuniform black string solutions
with horizon coordinate $r_0 =1$
and asymptotic length of the compact direction $L=L^{\rm crit}=7.1713$,
is shown for several values of the
nonuniformity parameter, $\lambda=0.5,~1,~2,~3,~5,~9$.
}
\\
\\
minimum radius ${\cal R}_{\rm min}$ representing the `waist'
are presented together with the 
proper length $L_{\rm H}$ of the horizon along the compact direction
as functions of the nonuniformity parameter $\lambda$,
ranging from $0 \le \lambda \le 9$.

With increasing $\lambda$, ${\cal R}_{\rm max}$ 
first increases, reaches a maximum around $\lambda \approx 4$
and then decreases slightly again towards still larger values
of the nonuniformity parameter;
in contrast $L_{\rm H}/L$ increases monotonically
and ${\cal R}_{\rm min}$ decreases monotonically (in the range
considered).
We expect that ${\cal R}_{\rm max}$ and $L_{\rm H}/L$
approach finite values in the limit $\lambda \rightarrow \infty$,
whereas ${\cal R}_{\rm min}$ should reach zero in this limit,
when extrapolated (approximately linearly) in the figure. 
For comparison, we also show in the figure the corresponding
geometric quantities of the $D=6$ nonuniform black string solutions.

\setlength{\unitlength}{1cm}
\begin{picture}(8,6)
\put(3,0.0){\epsfig{file=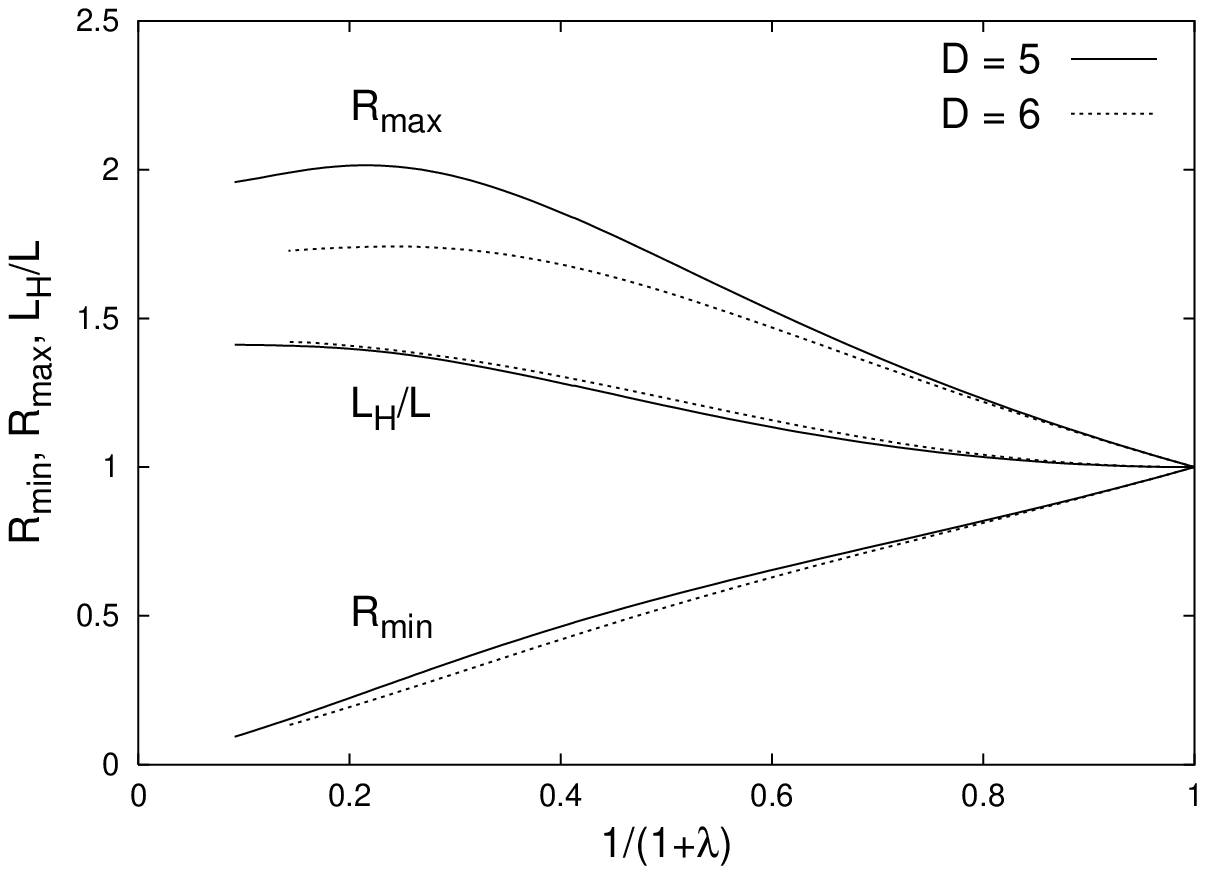,width=8cm}}
\end{picture}
\\
\\
{\small {\bf Figure 4.}
The maximum radius of the $(D-3)$-sphere on the horizon ${\cal R}_{\rm max}$,
the minimum radius ${\cal R}_{\rm min}$,
and the proper length $L_{\rm H}$ of the horizon along the compact direction
divided by the asymptotic length $L=L^{\rm crit}$
are shown for $D=5$ and $D=6$ nonuniform black string solutions
as functions of $1/(1+\lambda)$.
}
\vspace{0.5cm}
\\

We exhibit in Figure 5 the mass $M$, the relative tension $n$,
the temperature $T$ and the entropy $S$ of the $D=5$ 
and $D=6$ nonuniform string
solutions, in units of the corresponding uniform string solution,
versus the nonuniformity parameter $\lambda$. 
Interestingly, the mass and the entropy assume a maximal value
in the vicinity of $\lambda \approx 4$, 
while the tension and the temperature assume a minimal value,
both in 5 and in 6 dimensions. Since the extrema appear only
around $\lambda \approx 4$, 
which is the maximal value of $\lambda$
obtained in previous calculations \cite{Wiseman:2002ti},
they were not recognized there.
Extrapolating these quantities to $\lambda \rightarrow \infty$
yields for the tension the critical value $n_*$, where
$n_*/n_0 \approx 0.8$ for $D=5$ 
and $n_*/n_0 \approx 0.6$ in six spacetime dimensions.

The mass and tension exhibited in Figure 5 are obtained from the 1st law of thermodynamics
together with the Smarr relation (\ref{smarrform}).
A discussion of the mass and the string tension
as obtained from the asymptotic
fall-off of the metric functions is given in Appendix A.

\setlength{\unitlength}{1cm}
\begin{picture}(8,6)
\put(-1,0.0){\epsfig{file=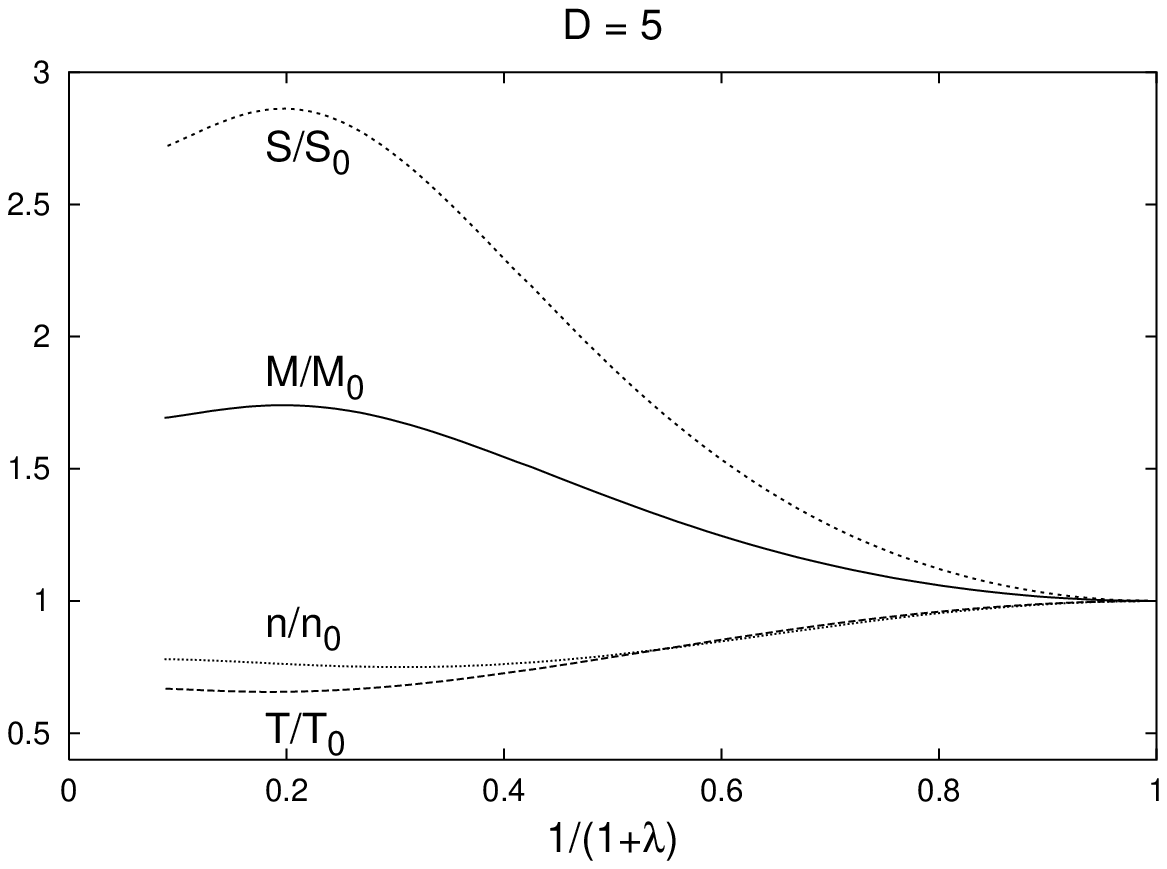,width=8cm}}
\put(7,0.0){\epsfig{file=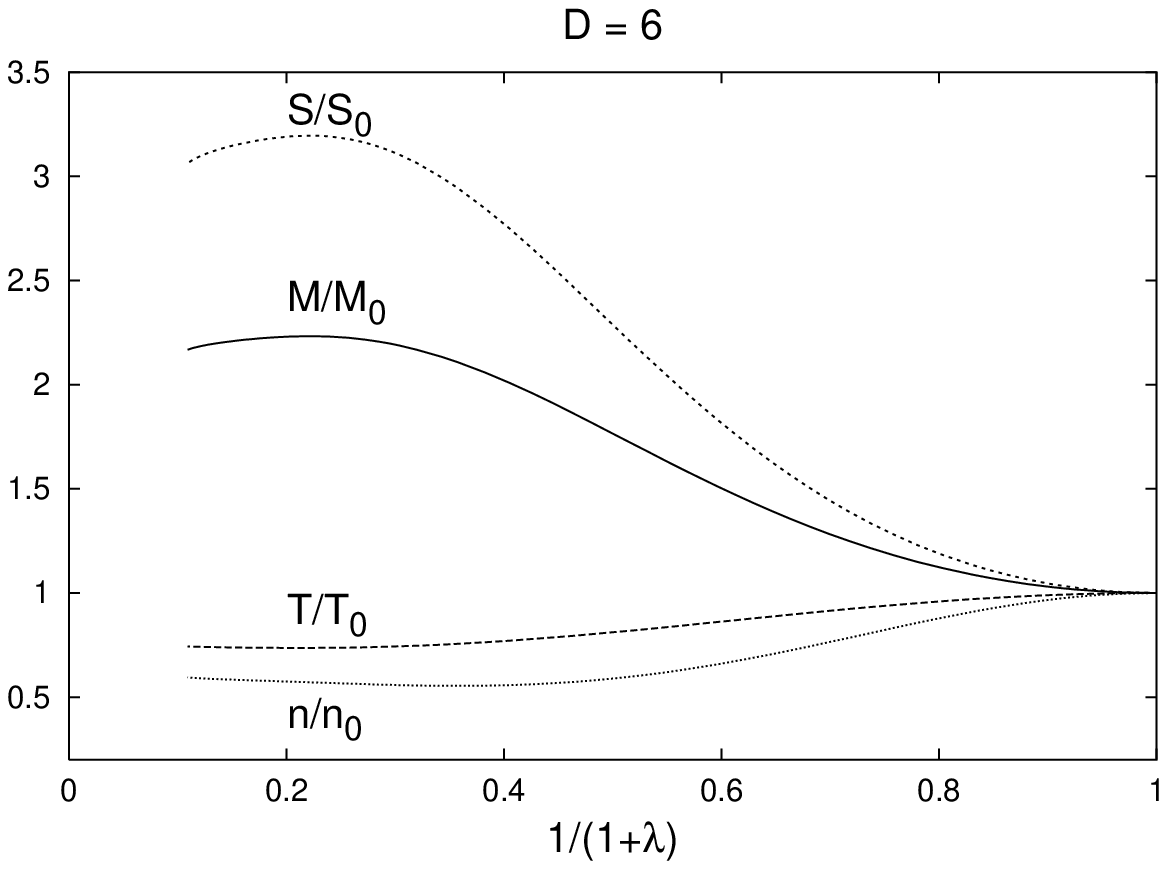,width=8cm}}
\end{picture}
\\
\\
{\small {\bf Figure 5.}
The $M$, the relative tension $n$, the temperature $T$ and the entropy $S$ 
of the $D=5$ (a) and $D=6$ (b) nonuniform string solutions
are shown
 in units of the corresponding uniform string solution
(denoted by $M_0$, $n_0$, $T_0$, and $S_0$) as functions of
$1/(1+\lambda)$.
}
\vspace{0.5cm}
\\

%%%%%%%%%%%%%%%%%%%%%%%%%%%%%%%%%%%%%%%%%%%%%%%%%%%%
\subsection{Black strings and black holes}
%%%%%%%%%%%%%%%%%%%%%%%%%%%%%%%%%%%%%%%%%%%%%%%%%%%%

In $D=6$ dimensions, evidence was provided that
the nonuniform string branch and the black hole branch merge at a 
topology changing solution \cite{Kudoh:2004hs}.
We would now like to reconsider this evidence in the light of
the continuation of the $D=6$ nonuniform string branch
to larger deformations,
and further address the question, whether there is analogous
evidence in $D=5$ dimensions.

We therefore exhibit in Figure 6 the mass $M$
versus the relative string tension $n$, 
for the nonuniform string branch as well as for the
black hole branch 
in 5 and 6 dimensions.
The black hole data are taken from \cite{Kudoh:2004hs}.
First of all we note qualitative agreement of the shape
and the relative position
of the nonuniform string branch and the black hole branch
in 5 dimensions with the shape and relative position
of the corresponding branches in 6 dimensions.
But compared to the data and discussion given in \cite{Kudoh:2004hs},
we here observe a new feature: the 
backbending of the nonuniform string
branch at a critical (minimal) value of the relative
string tension $n_b$.
Although the onset of this backbending can already be anticipated 
in the $D=6$ data of \cite{Kudoh:2004hs}.
But the backbending of the nonuniform string branch at $n_b$
is still in accordance with the conjecture of a
topology changing transition,
occurring at $n_* > n_b$, both in 5 and 6 dimensions.

We attribute the presence of the gap 
between the black hole branch and the nonuniform string branch
mainly to insufficient numerical data 
of the black hole branch close to the
anticipated transition point. At such a transition point
the nonuniform string parameter $R_{\rm min}$ must approach zero
(see Figure 4), and likewise the black hole parameter $L_{\rm axis}$
must approach zero, where $L_{\rm axis}$ measures the proper length along
the exposed symmetry axis \cite{Kudoh:2004hs}.

\setlength{\unitlength}{1cm}
\begin{picture}(8,6)
\put(-1,0){\epsfig{file=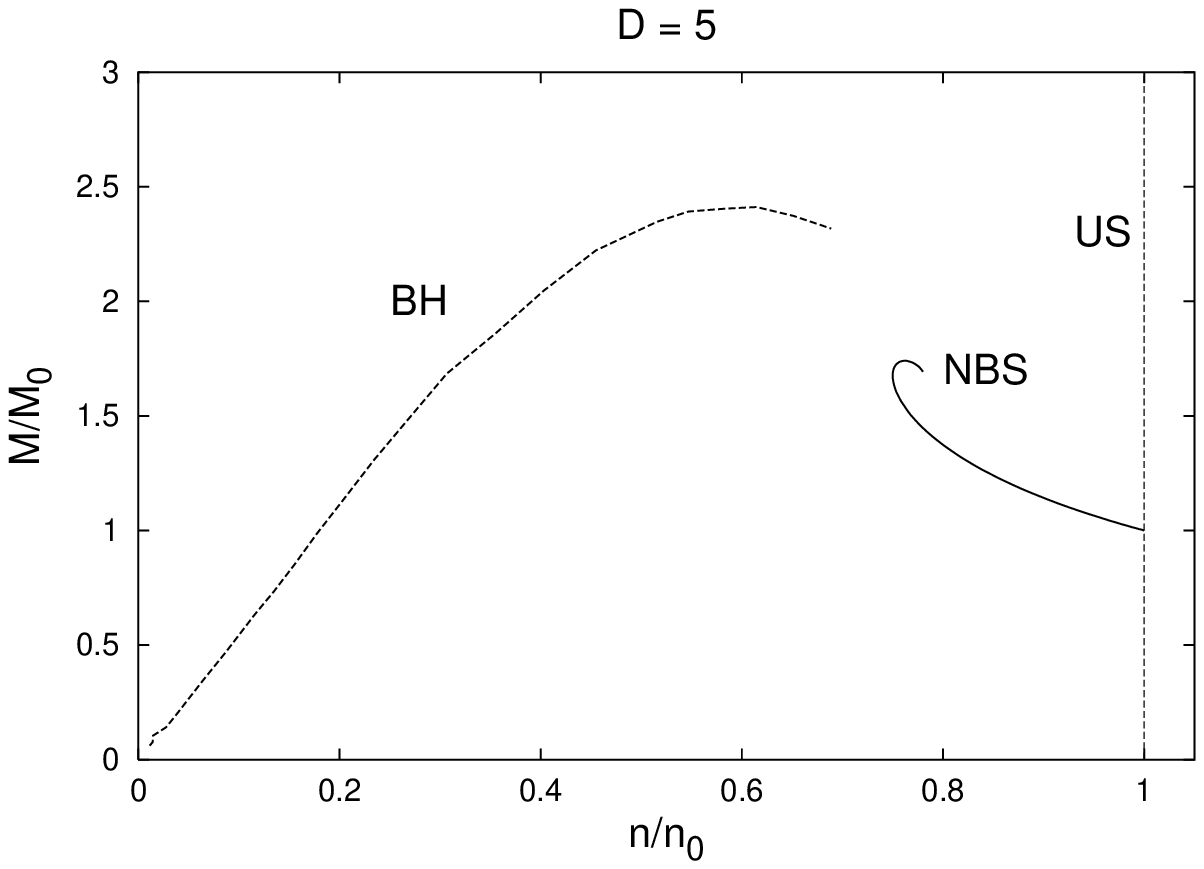,width=8cm}}
\put(7,0){\epsfig{file=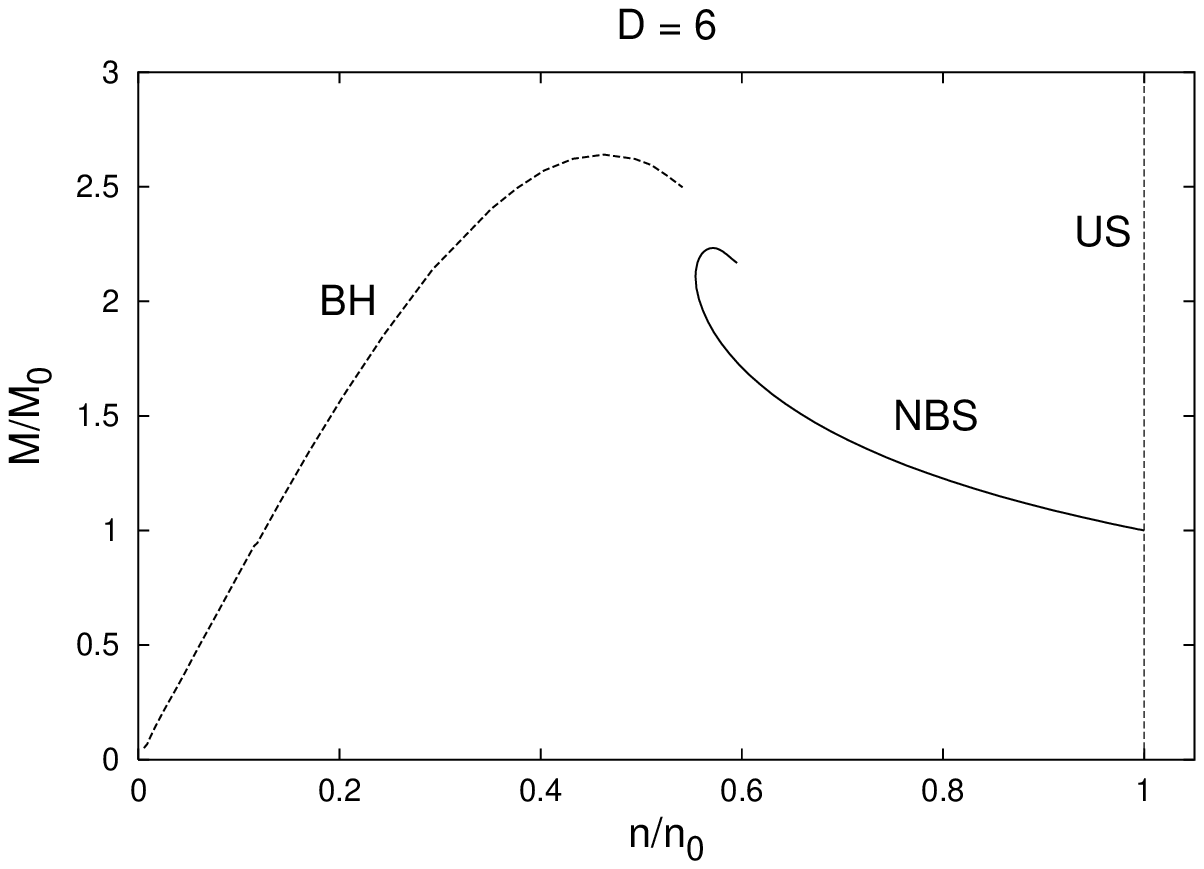,width=8cm}}
\end{picture}
\\
\\
{\small {\bf Figure 6.}
The mass $M$ of the $D=5$ (a) and $D=6$ (b) nonuniform string
and black hole branches
is shown versus the relative string tension $n$.
$M$ and $n$ are normalized by the values of the
corresponding uniform string solutions.
Here and in Figures 7-11, the data for the black hole branches is from
\cite{Kudoh:2004hs}.
}
\vspace{0.5cm}
\\
\\
We note, that
close to the anticipated transition point, $R_{\rm min}$
has decreased considerably farther 
(at the last numerically obtained point of the nonuniform string branch),
than $L_{\rm axis}$ (at the last numerically obtained 
point of the black hole branch).

Since $L_{\rm axis}$ has decreased farther for $D=6$ black holes
than for $D=5$ black holes, the gap between the branches
is smaller in 6 dimensions than in 5 dimensions.
We exhibit $R_{\rm min}$, $R_{\rm max}$, and $L_{\rm axis}$
as well as the black hole equatorial radius $R_{\rm eq}$
in Figure 7 for $D=5$ and $D=6$ solutions. 
$R_{\rm min}$ and $R_{\rm max}$
both exhibit the backbending feature present for
nonuniform string solutions at large deformations.
The figure is consistent with the vanishing 
of $L_{\rm axis}$ and $R_{\rm min}$ at the same critical value of $n$.
There $R_{\rm eq}$ and $R_{\rm max}$ should also merge.
The transition might then occur
in the vicinity of $n_*/n_0 \approx 0.8$ for $D=5$ and
$n_*/n_0 \approx 0.6$ for $D=6$
(as opposed to $n_{\rm b}/n_0 \approx 0.55$ for $D=6$, which was earlier
assumed to be the transition point \cite{Kudoh:2004hs}, but which is
now realized to be the point where the backbending occurs).
Extrapolating the black hole branch towards this critical value,
the $R_{\rm eq}$ curve appears to smoothly reach the endpoint of
the (backbending) upper part of the $R_{\rm max}$ curve
of the nonuniform string branch.
The black hole data are again taken from \cite{Kudoh:2004hs}.

Addressing the thermodynamic properties of the solutions,
we exhibit in Figure 8 and Figure 9 the temperature and the entropy
of $D=5$ and $D=6$ nonuniform strings and black holes.
Extrapolating the black hole branch towards the critical value $n_*$,
where the transition might occur,
the black hole curves for temperature and entropy
also appear to smoothly reach the endpoints of
the corresponding (backbending) upper parts of the nonuniform string branch.
This also holds for the mass, of course.
Again, the black hole data are from \cite{Kudoh:2004hs}.

\setlength{\unitlength}{1cm}
\begin{picture}(8,6)
\put(-1,0.0){\epsfig{file=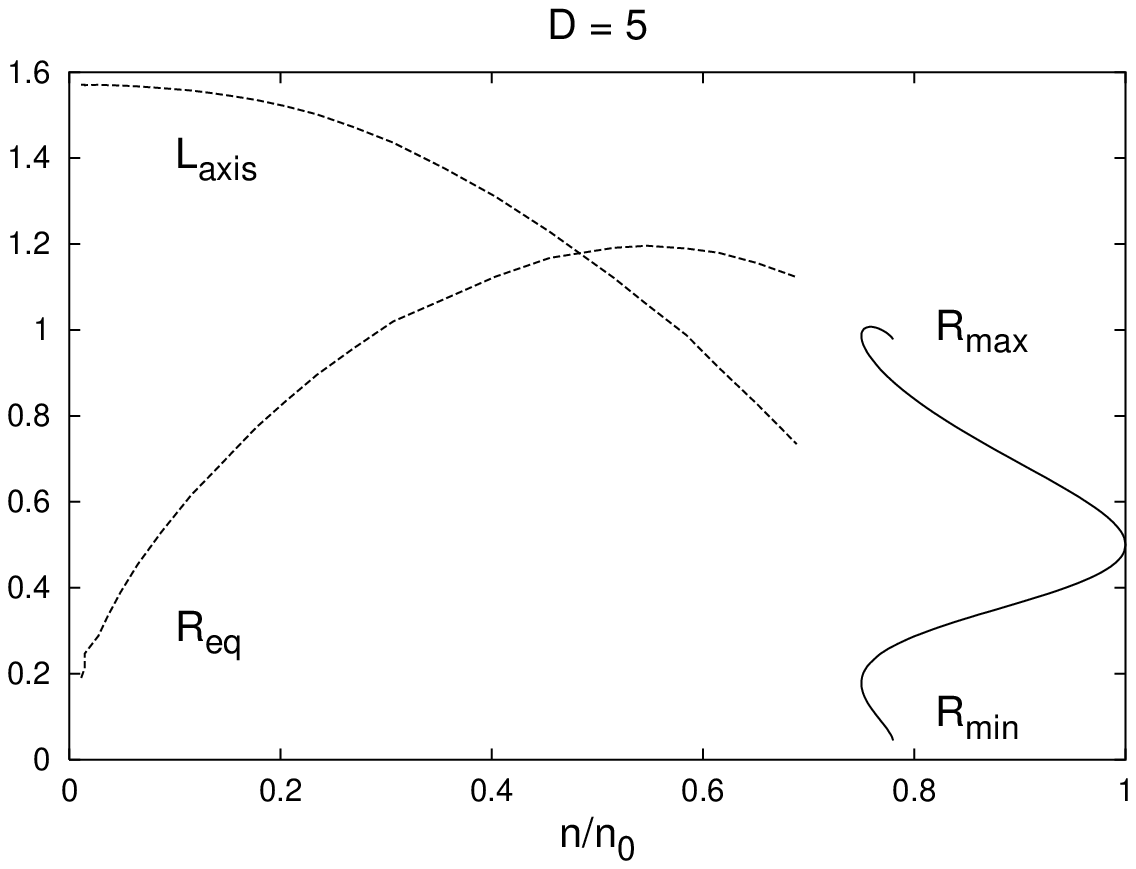,width=8cm}}
\put(7,0.0){\epsfig{file=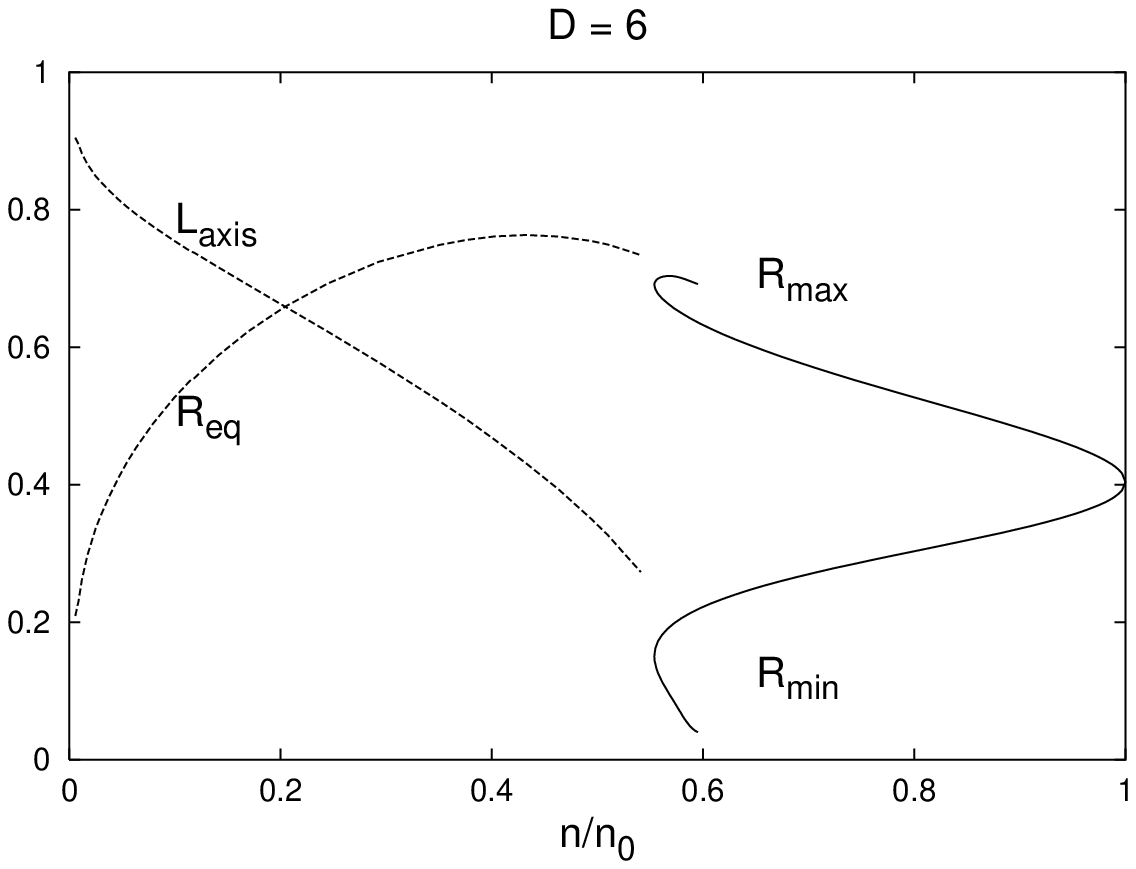,width=8cm}}
\end{picture}
\\
\\
{\small {\bf Figure 7.}
$R_{\rm min}$, $R_{\rm max}$, $L_{\rm axis}$ and $R_{\rm eq}$
of the $D=5$ (a) and $D=6$ (b) nonuniform string 
and black hole branches
are shown versus the relative string tension $n$. 
}
\vspace{0.5cm}
\\

For very small masses localized black holes are entropically favoured,
and for very large masses only uniform strings exist
\cite{Kudoh:2004hs}.
When the entropy is plotted versus the mass for the black hole branch
and the uniform and nonuniform string branches, one observes,
that the uniform strings become entropically favoured
at a certain value of the mass, lying above the critical string mass
and below the maximum black hole mass \cite{Kudoh:2004hs}.
This is illustrated in Figure 10 for solutions in five and six dimensions.
\\
\\
\setlength{\unitlength}{1cm}
\begin{picture}(8,6)
\put(-1,0.0){\epsfig{file=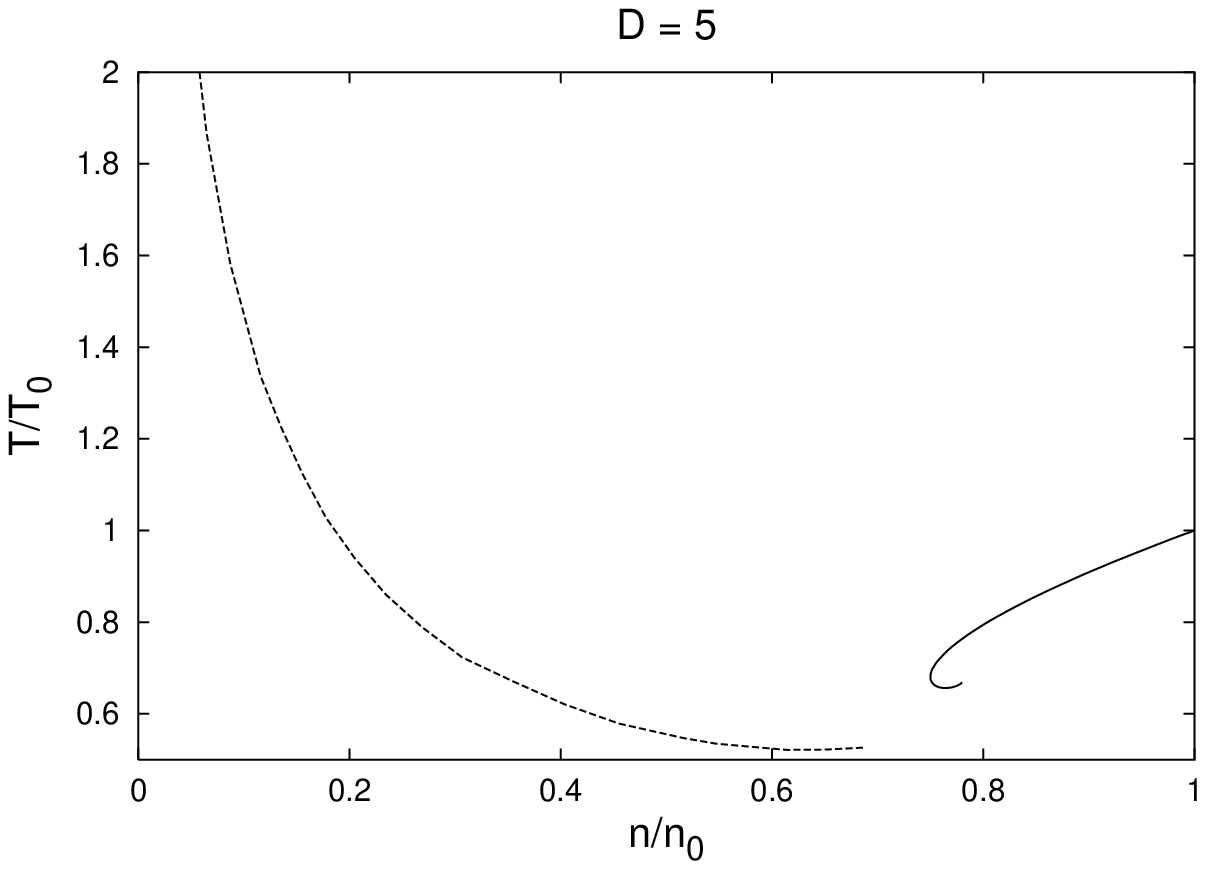,width=8cm}}
\put(7,0.0){\epsfig{file=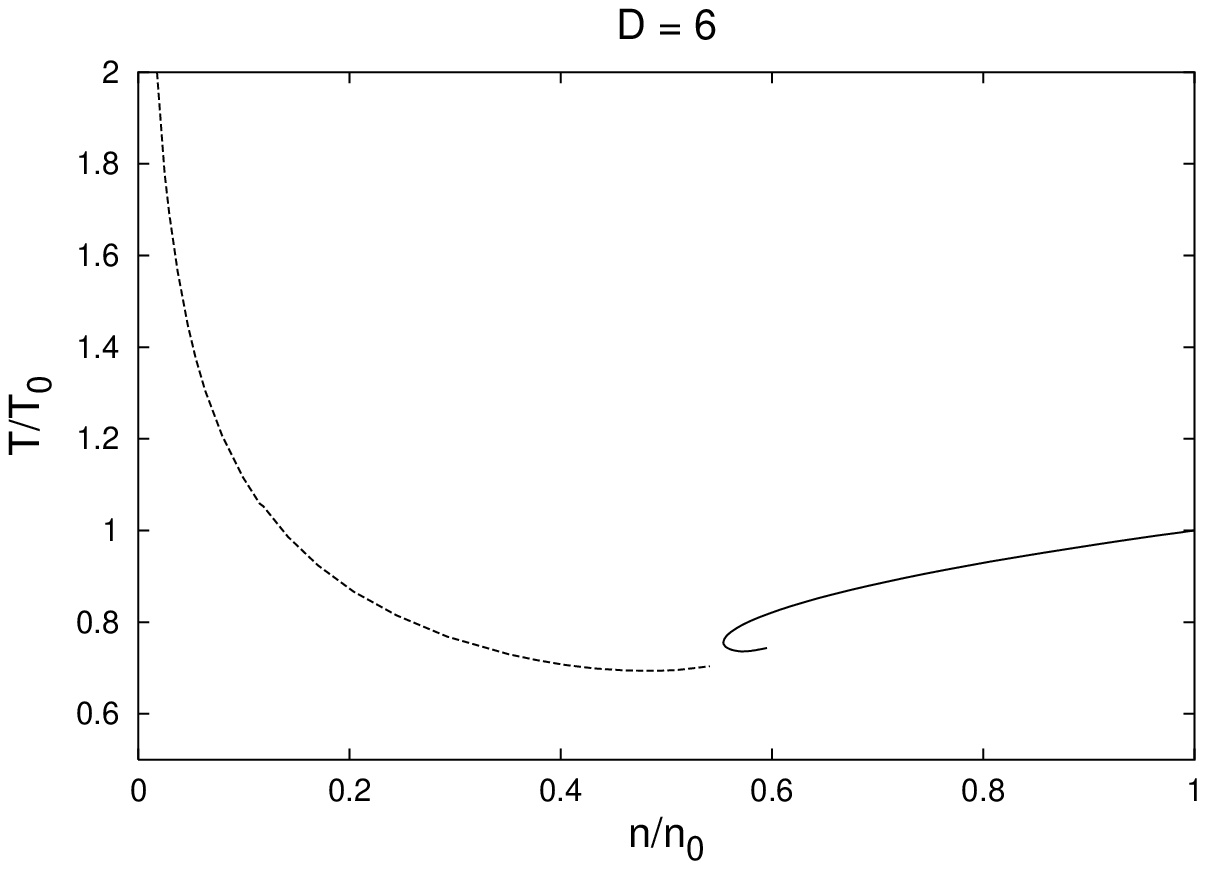,width=8cm}}
\end{picture}
\\
\\
{\small {\bf Figure 8.}
The temperature $T$ 
of the $D=5$ (a) and $D=6$ (b) nonuniform string
and black hole branches
are shown versus the relative string tension $n$
(in units of the uniform string quantities
$T_0$ and $n_0$). 
}
\vspace{0.5cm}
\\

%%%%%%%%%%%%%%%%%%%%%%%%%%%%%%%%%%%%%%%%%%%%%%%%%%%%%%%%%%%%%%%%%%%%%%%%%
\setlength{\unitlength}{1cm}
\begin{picture}(8,6)
\put(-1,0.0){\epsfig{file=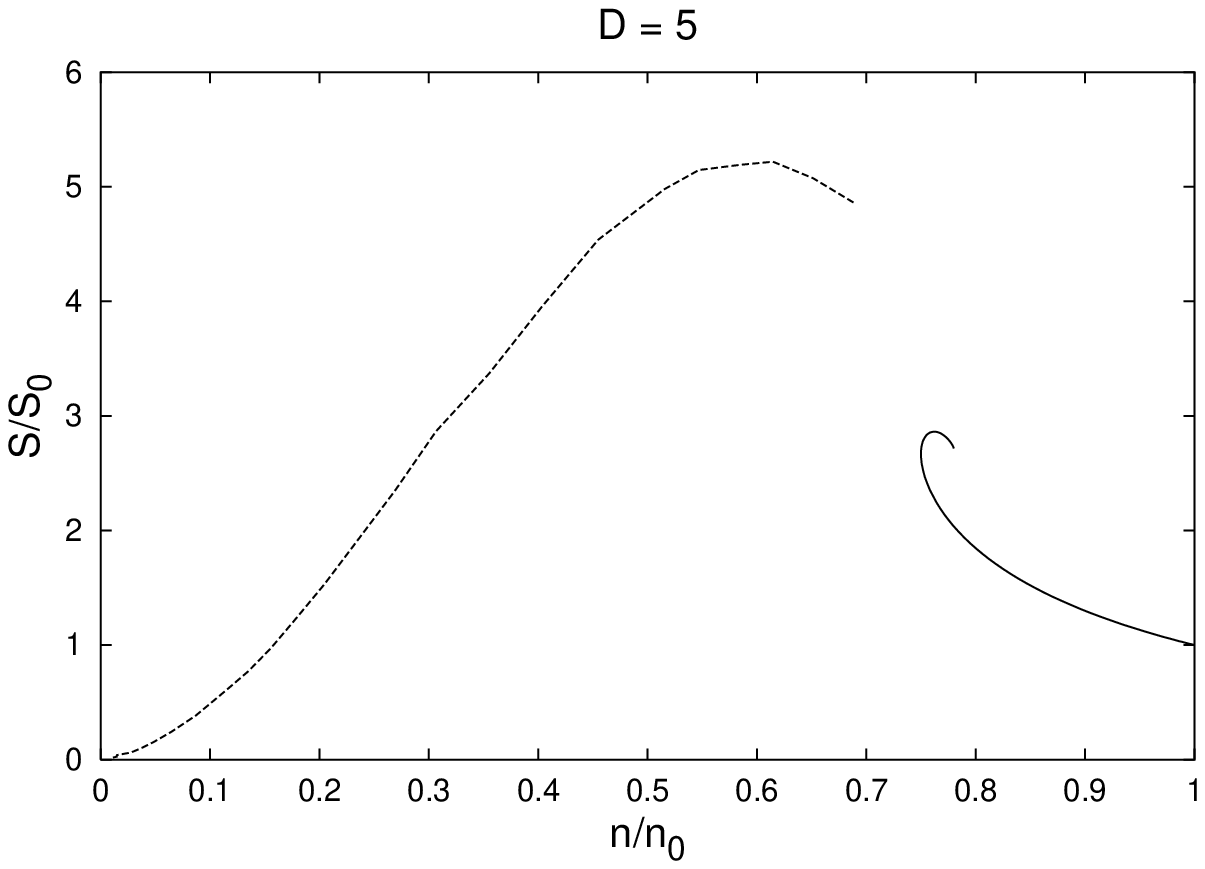,width=8cm}}
\put(7,0.0){\epsfig{file=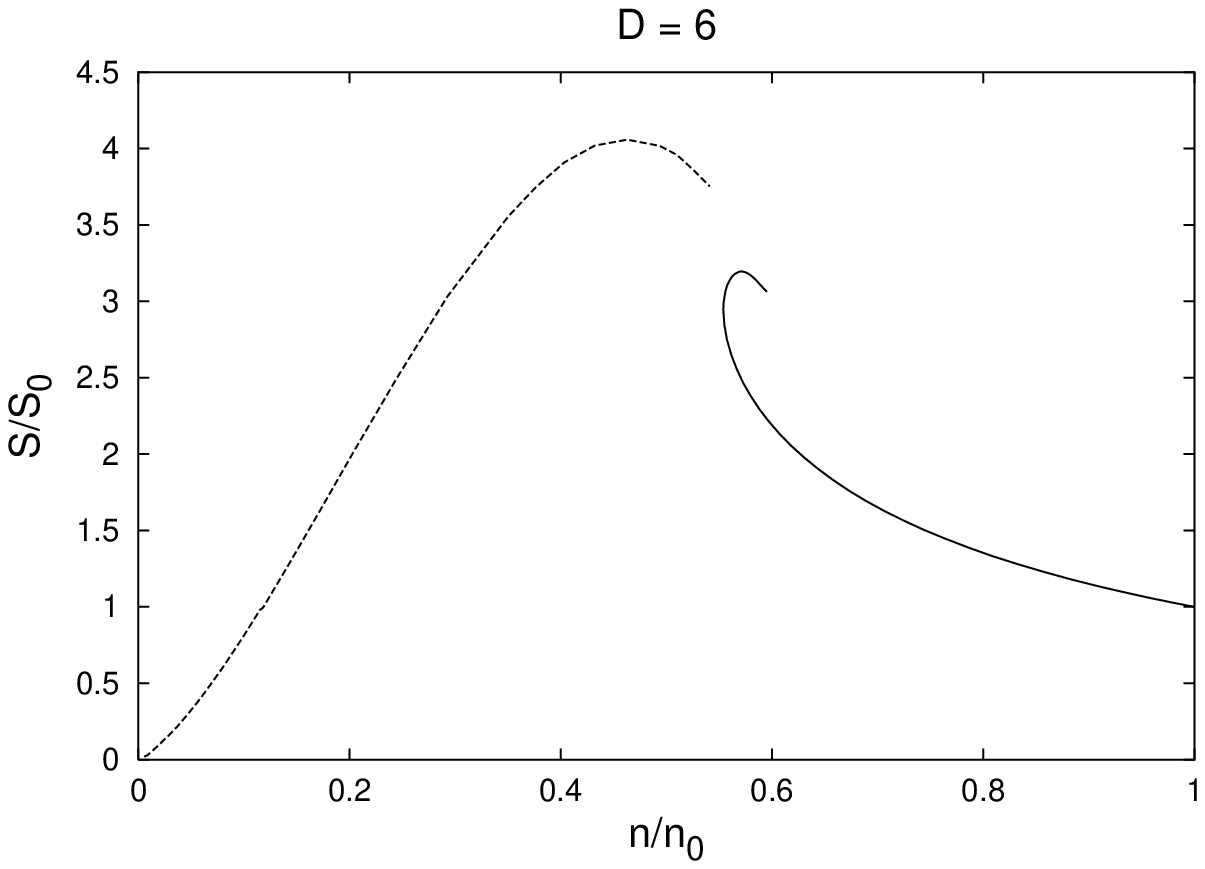,width=8cm}}
\end{picture}
\\
\\
{\small {\bf Figure 9.}
The the entropy $S$ 
of the $D=5$ (a) and $D=6$ (b) nonuniform string
and black hole branches
are shown versus the relative string tension $n$
(in units of the uniform string quantities
$S_0$ and $n_0$). 
}
\vspace{0.5cm}
\\
Concluding, we observe qualitative agreement of all the physical properties
of the solutions in 5 and in 6 dimensions.
This strongly suggests, that the same phenomenon
is present in both cases.
In particular, all data are consistent with the conjecture
that the black hole branch and the nonuniform string branch
merge in a topology changing transition. 
Our new nonuniform string solutions give further credence
to this scenario, but they still cannot confirm it.
\\
\\
\setlength{\unitlength}{1cm}
\begin{picture}(8,6)
\put(-1,0.0){\epsfig{file=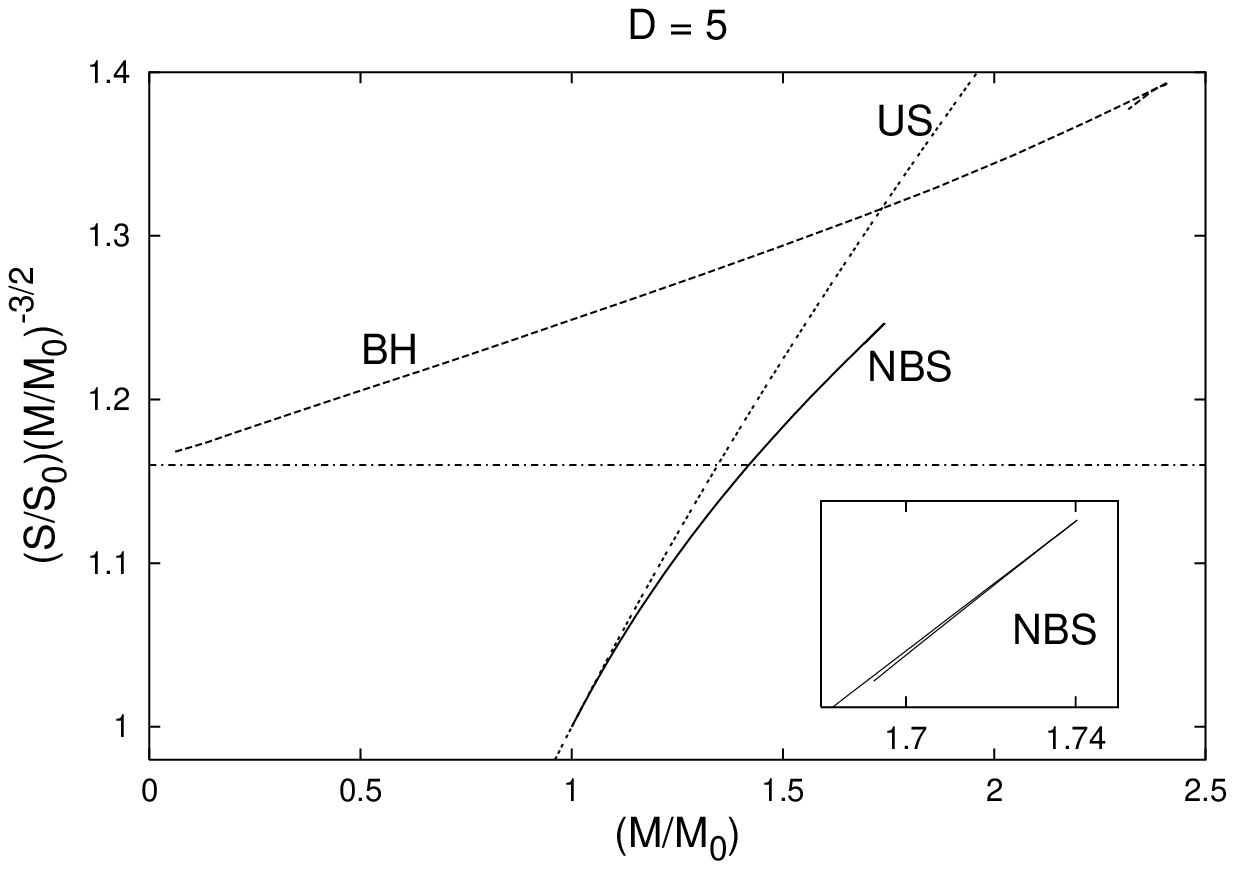,width=8cm}}
\put(7,0.0){\epsfig{file=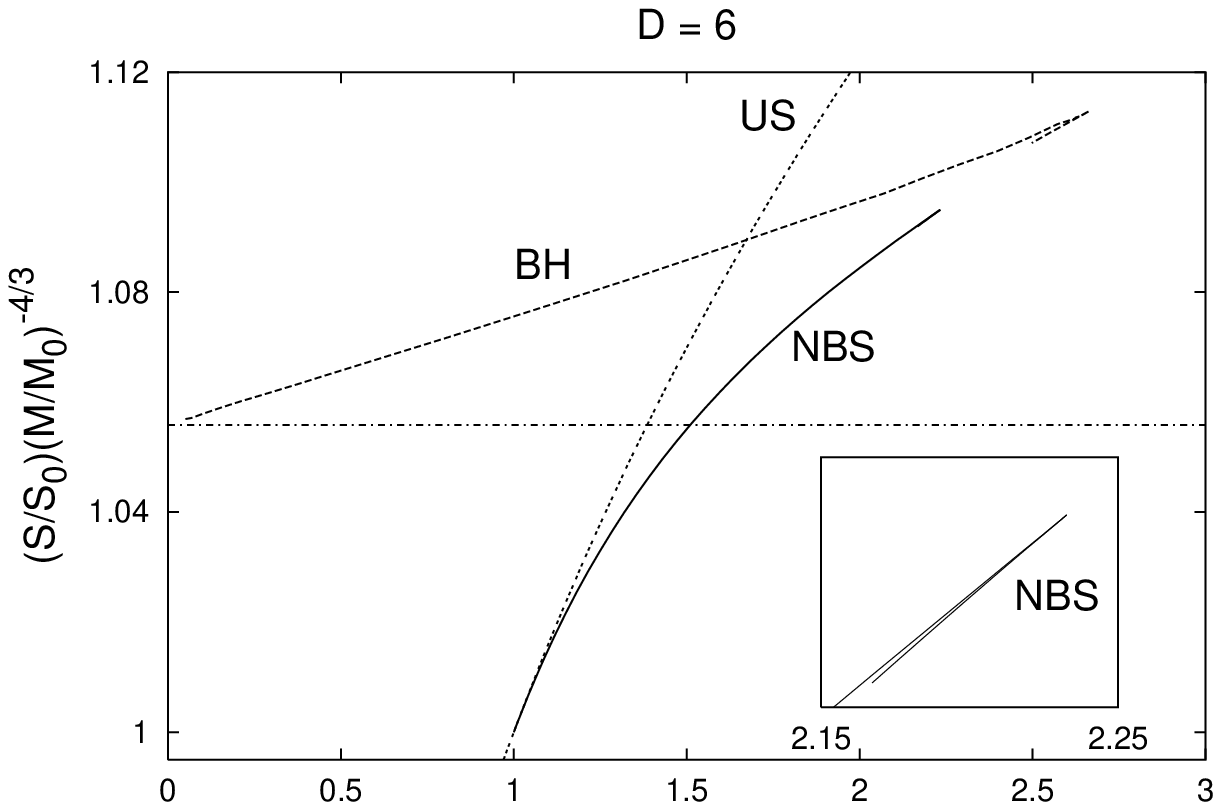,width=8cm}}
\end{picture}
\\
\\
{\small {\bf Figure 10.}
The product of entropy and mass $SM^{-3/2}$ (a) 
and $SM^{-4/3}$ (b) for the $D=5$ (a) and $D=6$ (b) 
nonuniform string branch,
the uniform string branch and the black hole branch
is shown as a function of the mass
(in units of the corresponding critical string quantities).
The horizontal lines represent the  curves
for the corresponding
Schwarzschild-Tangherlini solutions. 
}
\vspace{0.5cm}
\\

%%%%%%%%%%%%%%%%%%%%%%%%%%%%%%%%%%%%%%%%%%%%%%%%%%%%
\section{Black string solutions in Einstein-Maxwell-dilaton theory}
%%%%%%%%%%%%%%%%%%%%%%%%%%%%%%%%%%%%%%%%%%%%%%%%%%%%

We now consider the action describing
a gravitating Maxwell field coupled with a dilaton in a $D-$dimensional spacetime
\begin{eqnarray}
\label{actionEMd}
I=\frac{1}{16 \pi G}\int d^Dx \sqrt{-g} \left(R-
\frac{1}{2} g^{\mu \nu} \partial_\mu \phi \partial_\nu \phi
-\frac{1}{4} e^{-2a \phi} F^2  \right),
\end{eqnarray}
where $F=dA$.
The free parameter $a$ governs the strength of the coupling of the dilaton to the 
Maxwell field. 

The corresponding Einstein-Maxwell-dilaton (EMD) field equations are
\begin{eqnarray}
\label{eq1}
\nonumber
R_{\mu\nu} -\frac{1}{2}R g_{\mu \nu}= \frac{1}{2}~T_{\mu \nu},
\\
\label{eq2}
\nabla^2 \phi=-\frac{a}{2}e^{-2a\phi}F ^2,
\\
\nonumber
\label{eq3}
\partial_\mu (\sqrt{-g} e^{-2a\phi} F^{\mu \xi })=0.
\end{eqnarray}
with stress-energy tensor
\begin{eqnarray}
\nonumber
&&T_{\mu \nu}=T_{\mu \nu}^{(d)}+T_{\mu \nu}^{(em)},
\\
\label{Tik2}
&&T_{\mu \nu}^{(d)}=\partial_\mu \phi \partial_\nu \phi-\frac{1}{2}g_{\mu \nu}|\partial \phi|^2 ,
\\
\nonumber
&&T_{\mu \nu}^{(em)}=e^{-2a \phi}(F_{\mu \beta}F_{\nu \gamma}g^{\beta \gamma}
-\frac{1}{4}g_{\mu \nu}F^2).
\end{eqnarray}
Finding nonuniform black string and black hole 
solutions of these equations for a compact extradimension constitutes
a formidable technical task. However, it appears possible to 
use the 'hidden' symmetris of the model (\ref{actionEMd}) 
arising in the dimensionally reduced theory in order to
generate nontrivial
EMD solutions without actually solving the full set of equations.

This technique, known for the  $D=4$ Einstein-Maxwell theory since long,
was generalized in the last years to higher dimensions and various other matter fields.
In this paper, we'll follow the approach in \cite{Gal'tsov:1998yu},
supposing the existence of one 
Killing vector $\partial/\partial y$, and write the $D-$dimensional line element
in the form
\footnote{
Note that the Ref. \cite{Gal'tsov:1998yu} discusses a more general case, with an
action principle containing a $(d+1)$-differential form.
Therefore, the general metric ansatz 
(\ref{ansa1}) will contain a set of $d$-coordinates $y^i$. 
}
\begin{eqnarray}
\label{ansa1}
ds^2=g_{yy}(x)dy^2+h_{ij}(x)dx^i dx^j,
\end{eqnarray}
performing the KK reduction with respect to the $y-$direction.
As proven in \cite{Gal'tsov:1998yu}, the reduced action corresponds to
a non-linear $\sigma$-model, whose target space possesses a rich geometric structure.
These symmetries imply the existence of a Harrison transformation nontrivially acting on the  
spacetime variables and matter fields.
As a result, one may generate nontrivial solutions
of the $D-$dimensional EMD equations starting with 
known vacuum configurations.
A detailed description of this procedure and the explicit form of the Harrison
transformation is given in Ref. \cite{Gal'tsov:1998yu}
(see also the results in \cite{Yazadjiev:2005gs}).
{This approach is valid for any value
of the dilaton coupling constant (in particular
also for Einstein-Maxwell theory, i.e., for $a=0,~\phi=0$), 
and appears to be different from other results in literature.
Ref. \cite{Harmark:2004ws}, for example,
``charges up'' the neutral KK solutions
by uplifting them to eleven dimensional 
M-theory and employing boost and U-duality transformations, which fixes
a particular value of the dilaton coupling constant $a$.
The resulting solutions of type IIA/B string theory
describe non-extremal p-branes on a circle.
}

Here we present only the resulting solutions, which have 
rather different properties, depending whether $\partial/\partial y$
is a timelike or spacelike Killing vector.
Although the same generation techniques apply also for 
black hole solutions, we'll restrict to 
the black string case.

%%%%%%%%%%%%%%%%%%%%%%%%%%%%%%%%%%%%%%%%%%%%%%%%%%%%
\subsection{Asymptotically ${\cal M}^{D-1}\times S^1$ EMD black string solutions }
%%%%%%%%%%%%%%%%%%%%%%%%%%%%%%%%%%%%%%%%%%%%%%%%%%%%

We start with a vacuum black string solution, written in the following
form
\begin{eqnarray} 
\label{forma1}
ds^2=-V(x)dt^2+h_{ij}(x)dx^i dx^j.
\end{eqnarray}
where  $\partial/\partial t$ is a timelike Killing vector.  
 
The Harrison transformation in this case generates a
one parameter family of black string solution in EMD 
theory, with line element 
\begin{eqnarray} 
\label{forma2}
ds^2&=&-V(\cosh^2 \beta-\sinh^2 \beta V)^{-2\alpha (D-3)}dt^2+
(\cosh^2 \beta-\sinh^2 \beta V)^{2\alpha}h_{ij}dx^i dx^j,
\end{eqnarray}
and matter fields 
\begin{eqnarray} 
\label{forma3}
A_\mu&=&\sqrt{2(D-2)\alpha}
%\frac{\sinh \beta (\cosh \beta)^{1-2\alpha(D-3)} V}
\frac{\tanh \beta~e^{a\phi_0}V}
{\cosh^2 \beta-\sinh^2 \beta V }\delta_{\mu t},
\\
\nonumber
\phi&=&\phi_0-2a(D-2)\alpha\log(\cosh^2 \beta-\sinh^2 \beta V),
%~{\rm with~}
%\phi_0=4a\alpha(D-2)\log  (\cosh \beta),
\end{eqnarray}
where $\beta,~\phi_0$ are arbitrary real constants
and 
\begin{eqnarray}  
\label{alfa}
\alpha=(2a^2(D-2)+D-3)^{-1}.
\end{eqnarray}
For Einstein-Maxwell theory, we find $\alpha=1/(D-3)$, the corresponding value
in the Kaluza-Klein case being $\alpha=1/(2(D-2))$. 

Both uniform and nonuniform solutions of the EMD theory can be generated in
this way.
For example, the uniform string solution constructed 
within the metric ansatz (\ref{metric}) reads 
\begin{eqnarray} 
\nonumber
ds^2&=&-\frac{f(r)}{\left(1+(\frac{r_0}{r})^{D-4}\sinh^2 \beta\right)^{2\alpha(D-3)}}dt^2+
(1+(\frac{r_0}{r})^{D-4}\sinh^2 \beta)^{2\alpha}
 (\frac{dr^2}{f(r)}+dz^2+r^2d\Omega_{D-3}^2),
\\
A_\mu&=&\sqrt{2(D-2)\alpha}
%\frac{\sinh \beta (\cosh \beta)^{1-2\alpha(D-3)} V}
\frac{\tanh \beta~e^{a\phi_0}f(r)}
{1+(\frac{r_0}{r})^{D-4}\sinh^2 \beta}\delta_{\mu t},
\\
\nonumber
\phi&=&\phi_0-2a(D-2)\alpha\log(1+(\frac{r_0}{r})^{D-4}\sinh^2 \beta),
\end{eqnarray}
For all $a$, the surface $r=r_0$ is an event horizon, while
 $r=0$ is a curvature singularity.
The extremal limit is found by taking $\beta \to \infty$ together with
a rescaling of $r_0$ and has the form
\begin{eqnarray} 
\nonumber
ds^2&=&-(1+(\frac{c}{r})^{D-4})^{-2\alpha(D-3)}dt^2+
(1+(\frac{c}{r})^{D-4})^{2\alpha}
 (\frac{dr^2}{f(r)}+dz^2+r^2d\Omega_{D-3}^2),
\\
A_\mu&=&
\frac{\sqrt{2(D-2)\alpha}~e^{a\phi_0}}{1+(\frac{c}{r})^{D-4}}\delta_{\mu t},
~~~
\phi=\phi_0-2a(D-2)\alpha\log(1+(\frac{c}{r})^{D-4}),
\end{eqnarray}
$c$ being a real constant. 
Solutions describing several extremal black strings
do also exist \cite{Horowitz:2002dc}.

Returning to the general nonuniform string case,
we observe that its relevant properties  can be derived from the
corresponding $D-$dimensional vacuum solution.
The first thing to note   
is that the causal structure of the region 
$r>r_0$ is similar to the vacuum solutions;  in particular one finds the same
location of the event horizon.
For the metric ansatz (\ref{metric}), 
the spacetime  still approaches the
${\cal M}^{D-1}\times S^1$ background as $r \to \infty $, while the 
matter fields behave asymptotically as 
\begin{eqnarray}  
A_t\simeq \Phi+\frac{Q_e}{r^{D-3}},~~
\phi\simeq  \phi_0+\frac{Q_d}{r^{D-3}}~,
\end{eqnarray}
where $Q_e$ and $Q_d$ correspond, in a suitable normalization, 
 to the electric and the dilaton charges, 
respectively, $\Phi$ being the electrostatic
potential difference between the event horizon and infinity,
\begin{eqnarray}
\nonumber 
\Phi&=&\sqrt{2(D-2)\alpha}~e^{a\phi_0}\tanh \beta  ,
\\
Q_e&=&-\sqrt{2(D-2)\alpha}~e^{a\phi_0}\sinh \beta  \cosh \beta~c_t,
\\
\nonumber
Q_d&=&-2a \alpha (D-2)\sinh^2 \beta c_t.
\end{eqnarray}
The mass $\bar{M}$, the string tension $\bar{\mathcal T}$
and the relative string tension $\bar{n}$
of the EMD solutions are
\begin{eqnarray} 
\nonumber
\bar{M}&=&M(1+2(D-3-n)\alpha \sinh^2 \beta),
\\
\bar{\mathcal T}&=&{\mathcal T},
\\
\nonumber
\bar{n}&=&\frac{n}{1+2(D-3-n)\alpha \sinh^2 \beta}.
\end{eqnarray}
The electric charge and the dilaton charge can also
be expressed via (note also that the dilaton charge is not an independent quantity)
 \begin{eqnarray} 
Q_e&=&-\frac{M (D-3-n)}{ (D-4)}\sqrt{\frac{ \alpha }{ 2(D-2)}}
~e^{a\phi_0}\sinh 2\beta,
\\
 \nonumber
Q_d&=&-\frac{2a M}{D-4}(D-3-n)\alpha \sinh^2 \beta.
\end{eqnarray}
The relation between Hawking temperature $\bar{T}$
and the entropy $\bar{S}$ of the EMD solutions
and the corresponding quantities $T$ and $S$ of the vacuum seed solution is
\begin{eqnarray} 
\bar{T}=T(\cosh \beta)^{-2\alpha(D-2)},~~
\bar{S}=S(\cosh \beta)^{2\alpha(D-2)},
\end{eqnarray}
thus the product $TS$ remains invariant under the Harrison transformation. 

The Smarr relation (\ref{smarrform}),
derived in \cite{Harmark:2003dg} for the vacuum case,
admits a straightforward generalization to EMD theory,
\begin{eqnarray}
\label{smarrform1}
\frac{D-3-n}{D-2} \bar{M}= \bar{T}\bar{S} -\frac{(D-3)(D-4)}{D-2}\Phi \tilde{Q}_e,
\end{eqnarray}
where $\tilde{Q}_e=\Omega_{D-3}LQ_e$.
Therefore the thermodynamics of the EMD solutions can be derived 
from the vacuum solutions.
When the parameter $\beta$ is large one has a near extremal charged
black string.
However, a discussion of the extremal limit seems to require
knowledge of the region $r<r_0$ of the seed metric.

We conclude that every vacuum solution is associated
with a family of charged solutions, which depends on the parameter $\beta$.
In particular, the branch of non-uniform solutions emerging from
the uniform black string at the threshold unstable mode thus must
persist for strings with non-zero electric charge.  
The fact that the `phase diagram' of static solutions is qualitatively
unchanged as the charge varies strongly suggests that there is still an
instability for charged black strings \cite{Aharony:2004ig,Harmark:2005jk}.

Also, a discussion of the thermodynamical properties of these solutions appears possible.
This is interesting in connection with the Gubser-Mitra conjecture \cite{Gubser:2000ec}, 
that correlates
the dynamical and thermodynamical stability 
for systems with translational symmetry and infinite extent.
{This conjecture has been discussed by several authors in the last years 
(see e.g.~\cite{Harmark:2005jk}-\cite{Kudoh:2005hf}).
Ref.~\cite{Harmark:2005jk} uses the boost/duality transformation
to map the phases of KK solutions onto phases of 
non- and near-extremal Dp-branes with a circle in their transverse space.
The results there (see also \cite{Aharony:2004ig})
confirm the validity of the Gubser-Mitra conjecture
for non-extremal smeared branes.
Similar conclusions are found in 
Ref.~\cite{Kudoh:2005hf}, where   
a discussion of thermodynamical stability of charged black $p-$branes within
third-order perturbation theory is presented, 
the Gregory-Laflamme critical wavelength being also determined.
}

Asymptotically ${\cal M}^{D-1}\times S^1$ EMD black hole solutions 
can be generated by applying the same approach, starting with
vacuum seed solutions written in the form (\ref{forma1}).
The resulting solutions are still given by (\ref{forma2})-(\ref{forma3}),
with the corresponding expressions
of $V,~h_{ij}$.
Similar to the black string case, their properties
are completely determined by the vacuum seed black hole solutions,
the Smarr relation (\ref{smarrform1}) being also satisfied.

%%%%%%%%%%%%%%%%%%%%%%%%%%%%%%%%%%%%%%%%%%%%%%%%%%%%
\subsection{Black strings in a background magnetic field}
%%%%%%%%%%%%%%%%%%%%%%%%%%%%%%%%%%%%%%%%%%%%%%%%%%%%

A rather different picture is found in the case when  
 $\partial/\partial y$ in (\ref{ansa1})
is a spacelike Killing vector.
Here we start with a vacuum black string solution (\ref{metric})
written in the form 
\begin{eqnarray} 
ds^2= 
-e^{2A(r,z)}f(r) dt^2+e^{2B(r,z)}(\frac{dr^2}{f(r)}+dz^2)
+e^{2C(r,z)}r^2 (d \theta^2+
\\
\nonumber
\sin^2 \theta d \varphi^2+\cos^2 \theta d \Omega_{D-5}^2)
\end{eqnarray}
with $\theta \in [0,\pi/2]$, except for 
$D=5$ where $\theta \in [0,\pi]$.

After applying a Harrison transformation
to this configuration with respect to the Killing vector 
$\partial/\partial y\equiv \partial/\partial \varphi$,
one finds the following solution of the EMD equations
\begin{eqnarray} 
\label{M-metric}
ds^2&=&\Lambda^{2 \alpha}(r,z,\theta)
\Big(-e^{2A(r,z)}f(r) dt^2+e^{2B(r,z)}(\frac{dr^2}{f(r)}+dz^2)
\\
\nonumber
&&+e^{2C(r,z)}r^2 (d \theta^2 +\cos^2 \theta d \Omega_{D-5}^2)\Big)
+\frac{e^{2C(r,z)}r^2 \sin^2 \theta}{\Lambda^{2\alpha (D-3)}(r,z,\theta)}
d \varphi^2,
\\
A_{\mu}(r,z,\theta)&=&\frac{B_0 e^{2C(r,z)}r^2 \sin^2 \theta}{\Lambda(r,z,\theta)}\delta_{\mu\varphi},
\\
\phi(r,z,\theta)&=&-\frac{2a(D-2)}{2a^2(D-2)+D-3}\log \Lambda(r,z,\theta),
\end{eqnarray}
where
\begin{eqnarray} 
\Lambda(r,z,\theta)=1+ \frac{B_0^2(2a^2(D-2)+D-3)}{2(D-2)}e^{2C(r,z)}r^2\sin^2 \theta,
\end{eqnarray}
$B_0$ is a free parameter which characterizes the central strength of 
the magnetic field, and $\alpha$
is still given by (\ref{alfa}).
 
One can easily see that this solution  is, 
in terms of the usual definitions, a black string,
with an event horizon and trapped surfaces.
Again, the causal structure of the solution is not affected by the Harrison transformation.

To clarify it's asymptotics, we note that 
for $r\to \infty$, the line element (\ref{M-metric}) approaches the simpler form
\begin{eqnarray} 
\label{M-metric-back}
ds^2&=&\Lambda^{2 \alpha} 
(-dt^2+  d\rho^2 +dz^2+dz_1^2+\dots +dz_{D-4}^2) 
+ \Lambda^{-2\alpha(D-3)}\rho^2 d \varphi^2,
\\
\nonumber
\Lambda&=&1+ \frac{B_0^2 }{2\alpha(D-2)}\rho^2,
\end{eqnarray}
where we introduced the new coordinates $\{\rho,z_1,\dots z_{D-4}\}$ satisfying
$r\cos \theta=(z_1^2+\dots+ dz_{D-4}^2)^{1/2}$, $r \sin \theta=\rho$,
such that $[d(r \cos \theta)]^2+(r\cos \theta)^2 d \Omega_{D-5}^2=dz_1^2+\dots +dz_{D-4}^2$.

Therefore, asymptotically the solution (\ref{M-metric}) approaches the  
Melvin fluxbrane found in \cite{Gibbons:1987ps}, which
 represents a higher dimensional generalization of the four dimensional Melvin solution.
The Melvin magnetic universe is a regular and static, cylindrically symmetric solution
to Einstein-Maxwell(-dilaton) theory describing
a bundle of magnetic flux lines in
gravitational-magnetostatic equilibrium \cite{Melvin:1963qx}.
This solution has a number of interesting features,
providing the closest approximation in general relativity for a
uniform magnetic field.
There exists a fairly extensive literature on the properties of
this magnetic universe,
of particular interest being the black hole solutions in universes which are asymptotically Melvin
\cite{Hiscock:1980zf}.
Black hole solutions in a higher dimensional Melvin universe
have been constructed recently in \cite{Ortaggio:2004kr}.

Therefore, it is natural to interpret the EMD solution (\ref{M-metric}) 
as describing a black string in an external magnetic field.
Note that, similar to the asymptotically ${\cal M}^{D-1}\times S^1$ case, one 
may generate also KK black hole solutions in a Melvin fluxbrane 
background.

One can compute the mass and tension of the black string solutions by using the background
subtraction approach.
In this case, the natural background is the Melvin solution (\ref{M-metric-back}). 
It follows that, different from the case of
asymptotically ${\cal M}^{D-1}\times S^1$ solutions,
these quantities are still given by (\ref{2}).
 Moreover, it can be proven that the thermodynamic properties of the  
black string EMD solutions
are unaffected by the external magnetic field, $i.e.$ 
one finds the same expressions for the
Hawking temperature and entropy as in the $B_0=0$ no magnetic field case.
A similar property has been noticed in \cite{Radu:2002pn} 
for a $D=4$
Schwarzschild black hole in
a Melvin universe background (see also the $D>4$ extensions \cite{Ortaggio:2004kr}).
Therefore, this seems to be a generic property of uncharged black hole/black string
solutions in a background magnetic field extending to infinity.
Heuristically, this is due to the fact that, 
in the static case, the mass-point/string source of 
these configurations does not interact directly with the background magnetic field.

%%%%%%%%%%%%%%%%%%%%%%%%%%%%%%%%%%%%%%%%%%%%%%%%%%%%
\section{Conclusions}
%%%%%%%%%%%%%%%%%%%%%%%%%%%%%%%%%%%%%%%%%%%%%%%%%%%%

Our first concern has been the numerical construction
of $D=5$ nonuniform strings, representing solutions
of the vacuum Einstein equations.
While their physical properties
are similar to those of $D=6$ nonuniform
strings, their construction is more difficult.
We attribute this to the slower asymptotic fall-off
of the metric functions. 

The branch of nonuniform strings emerges smoothly from
the uniform string branch at the critical point,
where its stability changes \cite{Gregory:1993vy}.
Keeping the horizon coordinate $r_0$ and the asymptotic
length $L$ of the compact direction fixed,
the solutions depend on a single parameter,
specified via the boundary conditions.
Varying this parameter, the nonuniform strings
become increasingly deformed, 
as quantified by the nonuniformity
parameter $\lambda$. 
For the largest value of $\lambda$ reached,
the `waist' of the string
has a minimal radius of $R_{\rm min}\approx 0.1$ 
($R_{\rm min}\approx 0.13$)
for $\lambda=9$ ($\lambda=6$)
in $D=5$ ($D=6$) dimensions,
indicating that it is shrinking towards its
asymptotic value of zero when $\lambda \rightarrow \infty$.

Previously, in $D=6$ dimensions, evidence was provided 
that the nonuniform string branch and the black hole branch merge at a
topology changing transition \cite{Kudoh:2004hs}.
Although we see a backbending of the nonuniform string
branch in both $D=5$ and $D=6$ dimensions,
not observed previously, because the 
nonuniform string branch had not been continued to sufficiently
high deformation,
all our data are consistent with the assumption,
that the nonuniform string branch and the black hole branch
merge at such a topology changing transition.
In fact, extrapolation of the black hole branch towards
this transition point appears to
match well the (extrapolated) endpoint of the (backbending) part
of the nonuniform string branch.

For the phase diagram this would mean that we would have a region
$0 < n < n_{\rm b}$ with one branch of black hole solutions, then a region
$n_{\rm b} < n < n_*$ with one branch of black hole solutions 
and two branches of nonuniform string solutions, the ordinary one
and the backbending one,
and finally a region $n_* < n < n_0$ with only one branch of
nonuniform string solutions. 
(We here do not address the bubble-black hole sequences present for $n > n_0$).
Thus the topology changing transition would be associated with $n_*$,
and $n_{\rm b} < n < n_*$ would represent a middle region where
three phases would coexist, one black hole and two nonuniform strings.
This anticipated phase diagram is exhibited in Figure 11.

This is strongly reminiscent of the phase structure of the
rotating black ring--rotating black hole system in $D=5$ \cite{ER}.
The (asymptotically flat) rotating black holes have $S^3$ horizon topology,
and the (asymptotically flat)
rotating black rings have $S^2 \times S^1$ horizon topology.
The rotating black holes exist up to a maximal value of the
angular momentum (for a given mass), $0 < J < J_*$,
the rotating black rings are present only above a minimal value of the
angular momentum (for a given mass), $J_{\rm b} < J$,
and in the middle region $J_{\rm b} < J < J_*$ three phases coexist,
one black hole and two black rings \cite{ER}.

Further numerical work for nonuniform strings and in particular
for black holes 
in the critical region close to $n_*$
might confirm this picture further, and it might 
lead to further insight into the structure of the configuration space.
The backbending of the nonuniform string branch
clearly indicates that the configuration space is richer 
close to the anticipated

%%%%%%%%%%%%%%%%%%%%%%%%%%%%%%%%%%%%%%%%%%%%%%%%%%%%%%%%%%%%%%%%%%%%%%%%%
\setlength{\unitlength}{1cm}
\begin{picture}(8,6)
\put(3,0){\epsfig{file=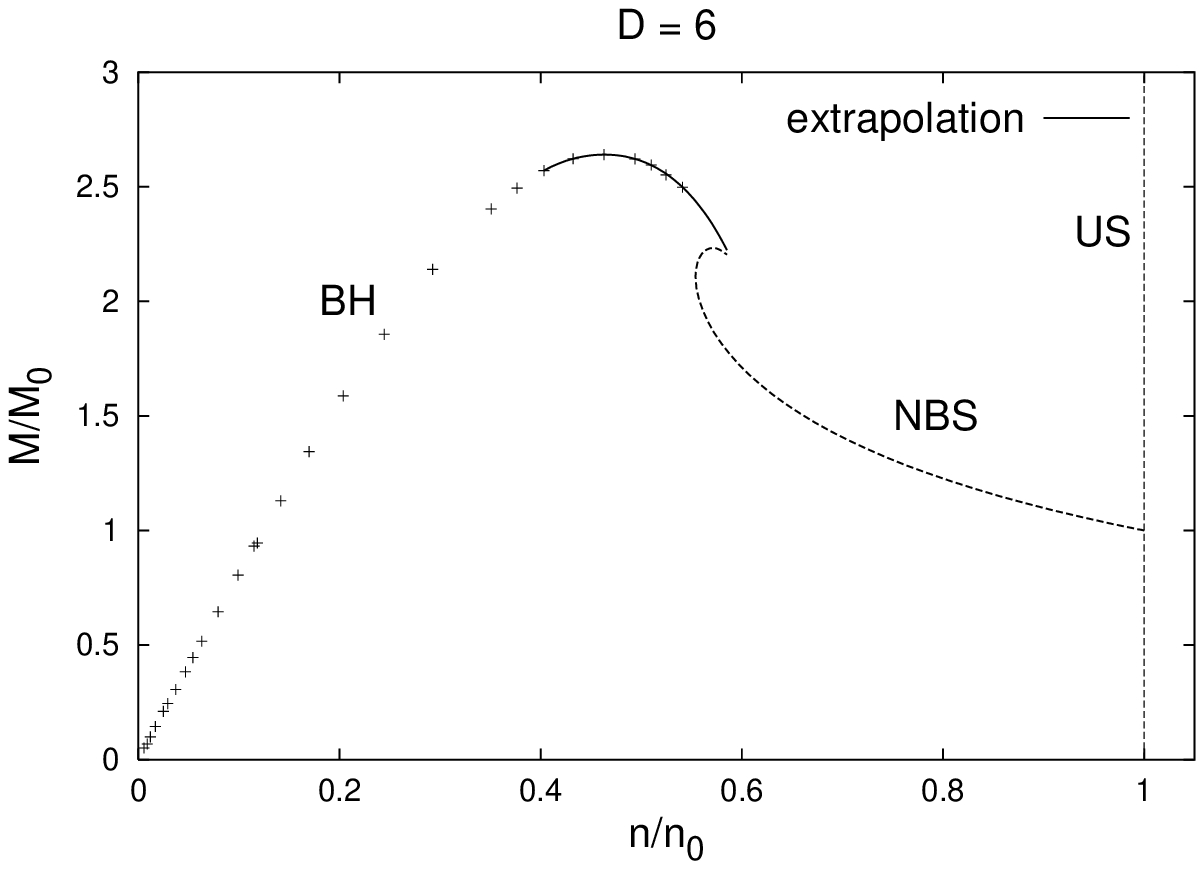,width=8cm}}
\end{picture}
\\
\\
{\small {\bf Figure 11.} 
The mass $M$ of the $D=6$ nonuniform string
and black hole branches
is shown versus the relative string tension $n$.
$M$ and $n$ are normalized by the values of the
corresponding uniform string solutions.
The black hole branch is extrapolated towards the
anticipated critical value $n_*$.
}
\vspace{0.5cm}
\\
\\
 topology changing transition,
which currently still invites speculations \cite{Kol:2004ww,foot1}.

Our second concern has been the construction
of black strings in EMD theory, obtained via a
Harrison transformation.
Different from other results in the literature (see e.g. \cite{Harmark:2004ws}),
the construction we proposed in Section 4 
is valid for any value of the dilaton coupling constant.
Apart from asymptotically 
${\cal M}^{D-1}\times S^1$ charged black strings,
we found also black string solutions in an external magnetic field.
We found that the properties of these EMD configurations can be derived from the
corresponding $D-$dimensional vacuum solution.

To push forward our understanding of these issues, it 
would be interesting to find both black hole and nonuniform 
black string branches for $D>6$.
The dimensions around $D=13$ are of particular interest, since
 as found in \cite{Sorkin:2004qq}, for $D>13$ the 
 perturbative nonuniform strings
 are less massive than the uniform solutions. Moreover, their entropy 
 is larger than the entropy of uniform strings with the same mass.

%%%%%%%%%%%%%%%%%%%%%%%%%%%%%%%%%%%%%%%%%%%%%%%%%%%%
\section*{Acknowledgements}
%%%%%%%%%%%%%%%%%%%%%%%%%%%%%%%%%%%%%%%%%%%%%%%%%%%%

We are grateful to T. Wiseman for valuable discussions
and for providing us with the $D=5$ and $D=6$ black hole data
\cite{Kudoh:2004hs}.
We would also like to thank D. H. Tchrakian for many useful discussions
and comments.
B.K.~gratefully acknowledges support by the DFG under contract
KU612/9-1.
The work of E.R. was carried out in the framework of Enterprise--Ireland
Basic Science Research Project SC/2003/390.
This work is also supported by IC/05/03.
\\
\\
%\newpage
% %%%%%%%%%%%%%%%%%%%%%%%%%%%%%%%%%%%%%%%%%%%%%%%%%%%%
\appendix
\section{Appendix}
\setcounter{equation}{0}

As noted before for black holes, 
the most error prone part of the numerical
procedure is the extraction of the mass $M$ and the
relative tension $n$ from the asymptotic form of
the metric via the coefficients $c_t$ and $c_z$
\cite{Wiseman:2002zc,Wiseman:2002ti,Kudoh:2004hs}.
Whereas this extraction is quite accurate for
nonuniform strings in 6 dimensions,
it is quite problematic for nonuniform strings in 5 dimensions.

To see this we consider the asymptotic expansion
for the metric functions in $D=5$ dimensions.
\begin{equation}
A \rightarrow \frac{A_\infty}{r} , \ \ \
B \rightarrow \frac{B_\infty}{r} , \ \ \
C \rightarrow \frac{C_\infty \ln r}{r} , 
\end{equation}
and in $D=6$ dimensions
\begin{equation}
A \rightarrow \frac{A_\infty}{r^2}  , \ \ \
B \rightarrow \frac{B_\infty}{r^2}  , \ \ \
C \rightarrow \frac{C_\infty}{r}   ,
\end{equation}
In $D=5$ the coefficients must satisfy $C_\infty=A_\infty+2 B_\infty$,
a relation essential for the first law to hold.

In 6 dimensions the asymptotic fall-off is sufficiently fast,
yielding excellent agreement for the mass and tension
as obtained from the expansion coefficients
with those obtained
from the first law and the Smarr relation
\cite{Wiseman:2002ti}-\cite{Kudoh:2004hs}.

In 5 dimensions the numerical situation
is much trickier, because of the presence of the
log term in the function $C$. This is aggravated
by the fact, that the coefficient $C_\infty$ is
an order of magnitude smaller than the other two coefficients.
We therefore do not see the log dependence in the
numerical results, when calculated in the full
interval $[0 \le \bar r \le 1]$, $[0 \le \bar z \le 1]$.
To observe the log dependence, 
we must switch to a system of ordinary differential equations
after a certain value of $\bar r$, beyond which 
the $\bar z$-dependence has disappeared.

The coefficient $c_t$ is nevertheless obtained with
good accuracy, since it is associated with a conservation law,
obtained from the equation for the metric function $A$ Eq.(\ref{Eeq}),
equivalent to the Smarr relation.
The coefficient $c_z$, in contrast, is error prone,
since the asymptotic fall-off of the function $B$
is numerically not well determined.
\\
\setlength{\unitlength}{1cm}
\begin{picture}(8,6)
\centering
\put(-0.5,0){\epsfig{file=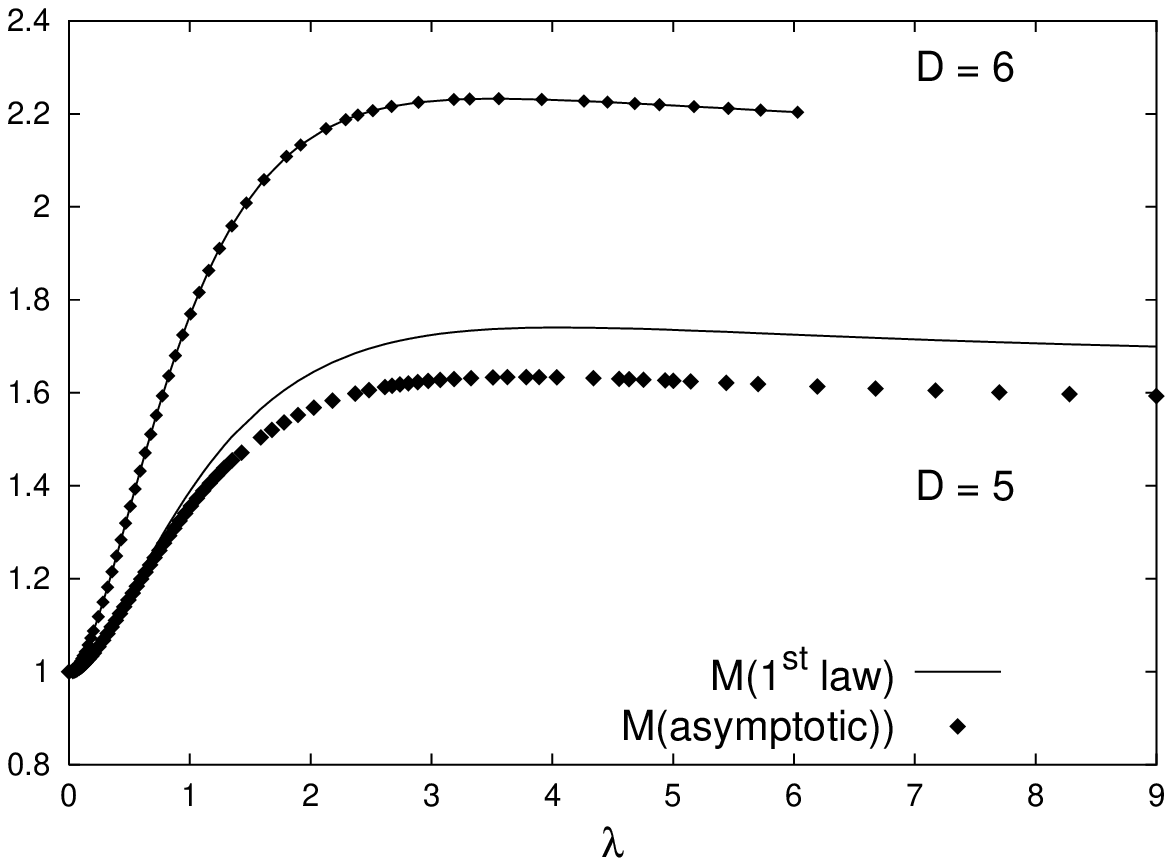,width=8cm}}
\put(7.5,0){\epsfig{file=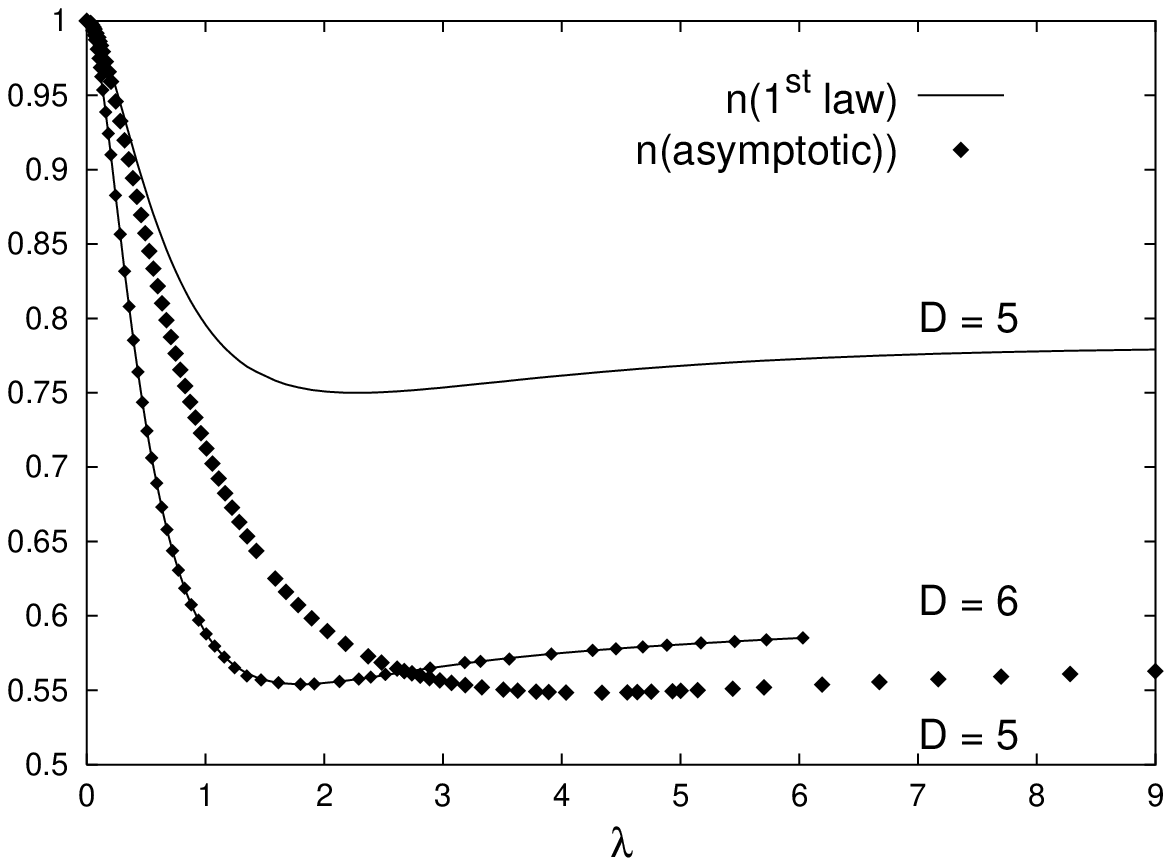,width=8cm}}
\end{picture}
\\
\\
{\small {\bf Figure 12.}
The mass $M$ (a)
and relative tension $n$ (b)
of the $D=5$ and $D=6$ nonuniform string branches
are shown versus
the nonuniformity parameter $\lambda$.
The quantities are extracted from the first law and the Smarr relation,
as well as from the asymptotic coefficients at $\bar r =1$.
}
\\
\\
Reading off the values of the coefficient $c_z$
at $\bar r =1$ therefore leads to considerable disagreement
of the values of the mass and tension
with those values obtained from the first law and the Smarr relation.
This is illustrated in Figure 12.

Our final remarks concern the numerical method and the quality of
the solutions.

For the construction of the numerical solutions we use 
Newton-Raphson iteration. In each iteration
step a correction to the initial guess configuration is 
computed. The maximum of the relative defect decreases 
by a factor of $20$ from one iteration step to another.
However, for large values of $\lambda$ convergence is slower.
In this case we re-iterate the solution until the defect is 
small enough (about $10^{-4}$). 
Note, that this defect concerns the discretised equations. 
The estimates of the relative
error of the solution (truncation error) are computed separately.
They are of the order $0.001$\% for small $\lambda$, but increase
up to $1$\% close to the backbending point. 
On the second branch the errors first decrease with increasing $\lambda$, 
but then increase again when the solutions become too steep at the waist.
The errors also depend on the order of consistency of the method,
i.e. on the order of the discretisation of derivatives. Usually $4^{\rm th}$ order
gives reasonable results.  
For some solutions we used $6^{\rm th}$ order to check the consistency with the 
$4^{\rm th}$ order solutions.

We also monitored the violation of the constraints
$$
      C_1 = \sqrt{f} \sqrt{-g} (G_r^r-G_z^z), \ \ \ 
                   C_2 =  \sqrt{-g} G_z^z \ .
$$		   
For small $\lambda$ the maximum of the constraints is less then $0.1$, 
but it increases with increasing $\lambda$. 
The maximum of the constraint $C_2$ is smaller by a factor of $5$
compared to $C_1$.  
On the second branch the maximum is of the order one. 
However, this large value appears in the vicinity of the waist 
($z=L/2$ at the horizon), in a small region where the functions  
are extremely steep. Away from this region the violation of the contraints remains small.
We also observed that the condition $\partial B/\partial r=0$ at the horizon 
is violated at the waist when $\lambda$ becomes large.
(A similar remark has been made in \cite{Wiseman:2002zc}.)
The violation of the condition $\partial B/\partial r=0$ at the waist might 
be related to the violation of the constraints.
       
Increasing the number of grid points near $z=L/2$ yields
smaller violation of the constraints and the conditon $\partial B/\partial r=0$ at 
the horizon.
Also, the violation of the constraint $C_1$ is of the same 
magnitude as the maximum of the equations with the same weighting.

%%%%%%%%%%%%%%%%%%%%%%%%%%%%%%%%%%%%%%%%%%%%%%%%%%%%%%%%%%%%%%%%%%%%%%%%%%%%%

\end{document}